\documentclass{nature}

\usepackage{graphicx}
\usepackage{multirow}
\graphicspath{ {images/} }
\DeclareGraphicsExtensions{.png,.pdf,.eps}

\usepackage{hhline}
\usepackage{epstopdf}
\usepackage{amsmath,amssymb}

\newcommand{\aj}{Astron. J.}

\newcommand{\apj}{Astrophys. J.}
\newcommand{\apjl}{Astrophys. J., Letters}
\newcommand{\apjs}{Astrophys. J., Suppl. Ser.}
\newcommand{\ao}{Applied Optics}

\newcommand{\aap}{Astron. Astrophys.}

\newcommand{\icarus}{Icarus}

\newcommand{\mnras}{Mon. Not. R. Astron. Soc.}

\newcommand{\pasp}{Publ. Astron. Soc. Pacific}

\newcommand{\nat}{Nature}

\newcommand{\jcp}{Journal of Chemical Physics}
\newcommand{\jgr}{Journal of Geophysics Research}
\newcommand{\jqsrt}{Journal of Quantitiative Spectroscopy and Radiative Transfer}

\usepackage{xcolor}

\newcommand*{\blue}{\textcolor{black}}

\title{The Impending Opacity Challenge in Exoplanet Atmospheric Characterization}

\author{Prajwal Niraula$^{1,*}$, Julien de Wit$^{1,*}$, Iouli E. Gordon$^{2}$, Robert J. Hargreaves$^{2}$, Clara Sousa-Silva$^{2}$, Roman V. Kochanov$^{2,3,4}$} 

\begin{document}
\maketitle

\textsl{$^*$These authors contributed equally to this work.}

\begin{affiliations}
 \item Department of Earth, Atmospheric and Planetary Sciences, Massachusetts Institute of Technology, 77 Massachusetts Avenue, Cambridge, Massachusetts 02139, USA;
 \item Harvard-Smithsonian Center for Astrophysics, Atomic and Molecular Physics Division, Cambridge, MA, USA;
 \item V.E. Zuev Institute of Atmospheric Optics, Laboratory of Theoretical Spectroscopy, Russian Academy of Sciences, 1 Akademichesky Avenue, 634055 Tomsk, Russia;
  \item QUAMER laboratory, Tomsk State University, 36 Lenin Avenue, 634050 Tomsk, Russia.
\end{affiliations}

\begin{abstract}
With a new generation of observatories coming online this decade, the process of characterizing exoplanet atmospheres will need to be reinvented. Currently mostly on the instrumental side, characterization bottlenecks will soon stand by the models used to translate spectra into atmospheric properties. Limitations stemming from our stellar\cite{rackham2018} and atmospheric\cite{macdonald2020} models have already been highlighted. Here, we show that the current limitations of the opacity models used to decode exoplanet spectra propagate into an accuracy wall at $\sim$0.5-1.0 dex (i.e., 3 to 10$\times$) on the atmospheric properties, which is an order of magnitude above the precision targeted by \textit{JWST} Cycle 1 programs and needed for, e.g., meaningful C/O-ratio constraints and biosignatures identification. We perform a sensitivity analysis using nine different opacity models and find that most of the retrievals produce harmonious fits owing to compensations in the form of $>$5$\sigma$ biases on the derived atmospheric parameters translating in the aforementioned accuracy wall. We suggest a two-tier approach to alleviate this problem involving a new retrieval procedure and guided improvements in opacity data, their standardization and optimal dissemination. 

\end{abstract}

\flushbottom
\maketitle

\thispagestyle{empty}
\section{Introduction}
Robust data interpretation requires building models that encompass the ensemble of physical processes possibly at play and accounting for the uncertainty associated with our understanding of these processes. For transmission spectroscopy, the four building blocks of such models relate to (1) the host star as a source of photons, (2) the planetary atmosphere attenuating the host star light, (3) the opacity describing the wavelength- and condition-dependence of the attenuation process, and (4) the instruments collecting the photons and converting them into electronic data. The associated uncertainty budget, long dominated by instruments through photon noise and systematics, needs a re-evaluation at the dawn of \textit{JWST} and the Extremely Large Telescopes (\textit{ELTs}).

A closer look at the current state of our models has already revealed challenges associated with the interpretation of the next-generation data\cite{burrows2014}. Stellar inhomogeneities can mimic or mute a planetary signal and our limited understanding of them may prove a bottleneck\cite{rackham2018}. The complex 3D structure of planetary atmospheres may play a similar role\cite{caldas2019}; simple 1D models are unable to capture the day-night temperature asymmetry\cite{macdonald2020} or highlight the role of  clouds and hazes in muting the spectral features\cite{betremieux2017, barstow2020}. Uncertainties (and biases) associated with atmospheric models are also expected to stem from our understanding--or lack thereof--of the underlying photochemical processes\cite{hu2012}. In addition, degeneracies amongst model parameters and dependencies to the model formulations will contribute to the challenges of robust high-precision retrievals\cite{heng2017, welbanks2019}.

The impact of uncertainties and inaccuracies in opacity models is a common hurdle. Atmospheric studies within the Solar System faced limitations associated with unavailable pressure broadening parameters\cite{dePater2005}. More recently, studies\cite{hedges2016, baudino2017, nezhad2019, barstow2020, greaves2021,ranjan2020} explored the possibilities for similar hurdles in the field of exoplanetary sciences and highlighted the large discrepancies between synthetic spectra generated from different pre-computed cross-sections supporting the need for additional experimental works\cite{fortney2019}. In 2021, a study detected atmospheric constituents of a hot-Jupiter via the cross-correlation technique using different sets of cross-sections, each yielding a different detection significance\cite{giacobbe2021}.

We expand the scope of these studies by assessing and quantifying the main sources of biases and uncertainties in opacity models considering their impact on the retrieved atmospheric parameters (scientific insights, i.e., final data product) rather than on the transmission spectra (observable, i.e., intermediate data product). It is a fundamental nuance as large differences between opacity models may not always translate into equivalent differences in the retrieved atmospheric parameters (see Methods). A transmission spectrum does not encode the atmospheric information uniformly across wavelengths\cite{batalha2017}. Its information density is a non-monotonic function of wavelength therefore mapping discrepancies in opacity models onto atmospheric parameters is non-trivial. To understand the impacts of imperfections in opacity models, we generate nine different sets of opacity cross-sections representative of standard assumptions and perform a perturbation/sensitivity analysis via self- and cross-retrievals (see Methods). 

\texttt{CS-DFLT} is our nominal cross-section generated using standard assumptions on the default calculation options of \texttt{hapi}\cite{kochanov2016} (v 1.0). \texttt{CS-1SUP} and \texttt{CS-1SDN} are designed to assess the effect of measurement uncertainties of line parameters from line lists to retrievals. \texttt{CS-SELF}, \texttt{CS-MAXB}, and \texttt{CS-MINB} are designed to assess the effect of incomplete knowledge regarding line broadening. \texttt{CS-500W} is designed to assess the effect of incomplete knowledge regarding far-wings of line profiles. While the aforementioned cross-sections use HITRAN2016\cite{hitran2016}, \texttt{CS-EXML} and \texttt{CS-HTMP} use respectively ExoMol\cite{exomol2020} and HITEMP2010\cite{HITEMP2010} to assess the effects of completeness and line-list imperfections through ``cross-database'' retrieval.

Figures~1 \&~2 shows the subtle differences emerging at the level of an opacity cross-section and a transmission spectrum, respectively, for warm-Jupiter observations simulated using each of the nine cross-sections. Although \texttt{CS-DFLT} (HITRAN) and \texttt{CS-HTMP} (HITEMP) are more similar to each other than to \texttt{CS-EXML} (ExoMol) at the level of cross-section for 0.1~atm due to their accounting of pressure shift (Figure~1.a), \texttt{CS-HTMP} and \texttt{CS-EXML} are more alike at the level of transmission spectra which probes mostly lower pressure levels where the effect of the pressure shifts is not as substantial but that of line-list completeness is (Figure~2). We present the transmission spectrum both in terms of wavelength-dependent variations of the transit depth and effective atmospheric height\cite{dewit2013}, $\rm h_{\rm eff}$, in scale height. Such a dual presentation of a transmission spectrum highlights the intermediate nature of this data product and connects it to the original signal and its noise (light-curve) and the final data product (atmospheric parameters).

\section{Results}
\subsection{Blind to opacity model perturbations.}
We find that despite the perturbations to the opacity models, most cross-retrievals lead to fits with reduced $\chi^2$ close to 1. Meaning that despite the opacity perturbations, it is possible to bias other parameters of the global model to compensate and yield a good fit (example in Figure~3). We find for example that perturbations to the opacity model through the broadening coefficient for our super-Earth scenario lead to biases on the atmospheric temperature, abundances of water, carbon dioxide, methane, and ozone of -4.02$\sigma$, 3.31$\sigma$, 6.91$\sigma$, 7.14$\sigma$, and 6.11$\sigma$, respectively.  We highlight that these biases affect each parameter individually. The abundances of some elements are overestimated while others are underestimated, which results in compounding biases on important elemental ratios, such as C/O for formation modeling. Additionally, we find that a cross-retrieval using the ExoMol and HITEMP2010 databases over HITRAN2016 for the same super-Earth scenario leads to similarly strong biases on the atmospheric temperature, abundances of water, carbon dioxide, methane, and ozone of -18.18/-19.37$\sigma$, -8.32/-20.64$\sigma$, -12.84/-16.42$\sigma$, -17.99/-65.95$\sigma$, and -8.61/-17.58$\sigma$ (see Methods), respectively.

We find that the significance of these biases is in some cases amplified by a shrinking of the posterior probability distributions (PPDs) of the derived parameters. For example, we find that the 1-D PPDs are 3 to 6 times narrower than in the nominal (i.e., self-retrieval) case. Such shrinking relates to a change in the $\chi^2$ distribution over the parameter space associated with a change in mapping from atmospheric parameters to transmission spectrum with different opacity models. We discuss further the complex topology of the $\chi^2$ distribution, the related relationship between the atmospheric parameter estimates, and the underlying dependency to the opacity models in Methods to provide further insights into the intricacies of this problem. This highlights that imperfections in opacity models may not only lead to biased parameter estimates, they may also affect the estimated uncertainty associated with them. Figure~4 highlights both effects. 

\subsection{Opacity-driven accuracy wall straight ahead.}
These biases are statistically significant and are also, and most importantly, physically substantial--often larger than 0.5 dex (equivalent to a factor of 3). While the former implies that our findings regarding opacity-induced biases are relevant within the context of observations reaching the signal-to-noise ratio necessary to highlight such biases, the latter is independent of such considerations and can be directly related to scientific insights that may thus be out of reach. Figure~5 highlights this difference between a statistically- and physically-significant bias while contextualizing the latter with the accuracy targeted by upcoming observation programs.

Figure~5.a shows the positive correlation between the Spectral Statistical Distance (SSD) and the Parametric Statistical Distance (PSD). The SSD is defined as $\sum  \sigma_\lambda^{-2} \times (\rm{Base~ Model}_{\lambda} - Model_{\lambda})^2$, where Base model is the synthetic spectrum calculated using the default cross-section (i.e. \texttt{CS-DFLT}), Model is the synthetic spectrum generated using another cross-section based on the same atmospheric parameters, and $\sigma_{\lambda}$ is the noise model associated with these observations (see Methods). The PSD is defined as the Bhattacharya distance\cite{bhattacharya1943} that measures the statistical distance between two PPDs while accounting for the covariance between the parameters. While SSD appears to correlate positively with PSD in log-log space, the large spread in value seen around their mean relationship highlights that perturbations at the level of a transmission spectrum cannot readily be translated into perturbations at the level of atmospheric parameters. Indeed, cross-sections initially yielding a large SSD may still enable a good fit (reduced $\chi^2\sim1$) owing to efficient compensation mechanisms (see Methods).

Figure~5.b introduces the same information as Figure~5.a, but without the context of a specific observation, defining the statistical nature of the PSD. Instead, Figure~5.b uses the Parametric Physical Distance, defined as $|1-\theta_{ret}/\theta_{true}|$, where $\theta_{ret}/\theta_{true}$ is the ratio of the retrieved versus true atmospheric parameter. Figure~5.b shows that, for most cross-retrievals, parameters are biased by $\sim$0.5 to $\sim$1.0 dex due to the opacity model. This corresponds to a 3 to 10$\times$ drop in accuracy compared to the accuracy expected from instrument models only for our synthetic observations. This 0.5-1.0 dex accuracy wall is best contextualized by the ambition of upcoming \textit{JWST} observations. GO~1633, for example, aims to constrain the metallicity of the hot Jupiter HD~189733~b to the 0.025-dex level to determine whether it is consistent with the Solar System’s relation. Such precision level is over one order of magnitude below the accuracy wall highlighted here. Our analysis reveals that for planetary systems similar to the synthetic systems used here (H-Mag~10, see Methods), this accuracy wall can be reached within 10 to 20 transits, while it would be reached within only a few transits for systems with bright hosts such as HD~189733~b (H-Mag~5.6). 

\section{Discussion}

We assess the extent to which complementary datasets such as emission spectra could help mitigate the observed biases by performing the same retrievals while introducing relevant priors on the p-T profile. We find that the biases were only marginally reduced (by up to 20\%). The current state of opacity models is thus yielding a hard accuracy wall around 0.5 to 1 dex. Such an accuracy wall will have severe consequences. For example, not being able to distinguish via transmission spectroscopy alone between a temperature of 300 K or 600 K, a “surface” pressure of 1 bar vs 2 bar, and/or an abundance of gases of 5\% vs 20\% will drastically impact our future capability to address key scientific questions spanning from formation mechanisms (e.g., via constraints on metallicity trends and elemental ratios) to the search for Life via the vetting of potential biosignature gases. 

\subsection{A two-tier mitigation strategy}
In order to support robust data interpretations and scientific claims while aiming to maximize the scientific output of \textit{JWST} and the \textit{ELTs}, we suggest the following two-tier approach. First, develop community standards for opacity models such as \texttt{tierraCrossSection} introduced in this work, to adequately estimate the sensitivity of derived inferences to opacity models.  This first step is aimed at propagating the biases and uncertainties from opacity models (line lists, cross-sections, and thermodynamic data) to planetary parameters. Doing so will provide robustness to scientific inferences but will also substantially limit the scientific outputs of the aforementioned observatories (marked as ``Accuracy Drop'' in Figure~5.b). For that reason, the second step suggested aims at investigating the origins of the biases and uncertainties in current opacity models that dominate biases and uncertainties on scientific inferences to guide the improvement of these models and match the accuracy enhancement within reach with \textit{JWST} and the \textit{ELTs} (marked as dotted line in  Figure~5.b).

\subsection{Identifying the dominant bottlenecks of opacity models}
We find that the effect of measurement uncertainties (\texttt{CS-1SUP} and \texttt{CS-1SDN}) leads to the least significant biases (see Figure~5). The biases found are contained below 0.15-dex and 4$\sigma$. We find that this effect is currently dominated by uncertainties on the line intensity and to a lesser extent the line precision, which will be a bottleneck for future missions (see Methods). Accounting for the fact that these two cross-sections were generated to represent the maximal compounding effect of the line list uncertainties, we thus find that the current precision at which line parameters are reported is adequate for a robust interpretation of \textit{JWST} data--barring the absence of systematic errors and the completeness of the list (see \texttt{CS-EXML} and \texttt{CS-HTMP}). 

We find that the effect of incomplete knowledge regarding line broadening (\texttt{CS-SELF}, \texttt{CS-MAXB}, and \texttt{CS-MINB}) leads to strong biases reaching values of up to 1-dex and 7$\sigma$. We expect that the additional consideration of collisional broadening would only further compound the underlying problem.   

We find that the effect of incomplete knowledge regarding far-wings of line profiles (\texttt{CS-500W}) is larger for the super-Earth synthetic case than for the warm-Jupiter one (see Figure~5). While for the latter the biases are confined below 0.5-dex and for all but one parameter, for the former the biases are around 1-dex for half of them. This difference is explained by the high sensitivity of the retrieved abundances of molecules with strong parse absorption features such as methane to their line profiles, combined with their connection to the atmospheric mean molecular mass and thus scale height--which is independently constrained by the data\cite{dewit2013}--for which all the other atmospheric parameters (incl. abundances and temperature) can try to compensate (see Methods).

We find that ``cross-database'' retrievals (\texttt{CS-EXML} and \texttt{CS-HTMP}) produce the strongest biases due to completeness issue, as expected (Figure~2). Molecules known to have discrepant line lists such as methane\cite{hargreaves2020} understandably have the strongest biases -17.99/-65.95$\sigma$ and -11.2/-20.9$\sigma$ for super-Earth and warm-Jupiter, respectively. Although the significance of these biases is increased by reducing PPDs (by a factor up to 6, Figure~4), they still correspond to biases up to 2~dex. Fortunately, we find that the residuals of such retrievals had reduced $\chi^2$ significantly above one (see Methods), which indicates a strong mismatch between data and model. This is however the only case where the model parameters could not allow for total compensation of the cross-section perturbation. 

\subsection{Steps towards alleviating the dominant bottlenecks of opacity models}

Following these findings, we suggest that the accuracy wall stemming from opacity models would be best addressed via a refined understanding and parametrization of (1) line broadening effects and (2) far-wing behaviors together with (3) a refined tracking of differences between line lists and underlying data sources. (1) and (2) would benefit from additional laboratory measurements and theoretical calculations.

Regarding (1) the line broadening effect specifically, measurements and calculations of collisional broadening for any relevant species, as well as an analysis of the temperature dependence of these parameters are needed. Currently, there are significant gaps in the knowledge of broadening behavior of many molecules of interest for atmospheric characterization. Traditionally, the majority of the efforts in this area have focused primarily on spectral lines for terrestrial atmospheric molecules broadened by air. Recently, the HITRAN database started to add broadening parameters due to the pressure of other ambient gases dominating planetary atmospheres, including H$_2$, He, CO$_2$ (see for instance ref.\cite{10.1016/j.jqsrt.2015.09.003}) and H$_2$O\cite{10.1029/2019JD030929}. As described in ref.\cite{10.1016/j.jqsrt.2015.09.003} there was a severe lack of laboratory measurements or calculations for some of these molecules, and crude estimates had to be used. The call for more measurements has been heard for some of the systems. For instance, only one line of SO$_2$ broadened by CO$_2$ had been measured\cite{Krishnaji1963} prior to the publication of ref\cite{10.1016/j.jqsrt.2015.09.003}. Since then, however,  several experimental and theoretical works have appeared\cite{Ceselin2017,Dudaryonok2018,10.1016/j.jqsrt.2018.12.030}. Nevertheless, the situation has not improved for many other molecules, for instance, dozens of articles are devoted to measurements of the CO$_2$ lines broadened by CO$_2$, O$_2$ and N$_2$\cite{10.1016/j.jqsrt.2020.107283}, but almost none for CO$_2$ broadened by H$_2$. 

It is worth noting that experiments that measure broadening parameters are not trivial to perform. For instance, ref\cite{Gordon2007} shows examples when a single parameter was measured in different laboratories but the reported values often did not agree within their uncertainties. However, with the advent of modern experimental techniques and fitting softwares, the situation is improving. The quality of the theoretical calculations of the broadening parameters has also improved recently\cite{Nguyen2020, Jozwiak2021}, although these are still quite limited since the calculations are computationally expensive, especially when carried out for a wide range of temperatures. Nevertheless, theoretical calculations do provide a promising avenue to complement experimental measurements.  

Regarding (2) the far-wing effect specifically, this relates to  one of the least studied and understood areas of molecular spectroscopy. For different molecules and their possible collisional broadeners (or perturbers), the difference in far-wing behavior can be drastic\cite{hartmann2002}. As an example, the far wings of water vapor lines, which are major contributors to the water continuum\cite{Mlawer2012}, are parametrized differently than those of CO$_2$ lines that are assumed to be sub-Lorentzian\cite{Cousin1985}.

Regarding (3) the refined tracking of differences between line lists and underlying data sources, the community would benefit from evaluations between theoretical line lists and  available laboratory data\cite{hargreaves2020}, and atmospheric/planetary data\cite{Toon2016, Olsen2019}. For several molecules, there are multiple sources of opacity data, each with different formats, advantages, and limitations. A centralized database, with standardized formats and informative metadata, is a crucial step in responsibly providing reliable and up-to-date spectral data to the astronomical community.

Each of the aforementioned actions are key to alleviating the opacity-driven accuracy wall revealed here, which is likely to similarly influence other remote sensing techniques and applications. Therefore, such additional work should be viewed as foundational to the success of  \textit{JWST} and the \textit{ELTs}, and a priority for our community.

\vspace{0.5cm}

\newpage
\pagebreak
\clearpage

\begin{methods}

\subsection{Framework}

We map the key components to a transmission spectrum to guide the development of our framework aimed at assessing the effect of imperfections in opacity models on retrieval capabilities. Supplementary Figure \ref{fig:framework} shows these components, and our process to identify accuracy bottlenecks. The framework uses minimal complexity for stellar, atmospheric, and instrumental models to focus on the role of opacity models. We notably assume long observations to reach a regime that allows us to reveal from the observation noise the limitations stemming from opacity models. Doing so allows us to study how biases and uncertainties propagate from line lists to transmission spectra and, ultimately, retrieved atmospheric parameters via a perturbation/sensitivity analysis. We perform such sensitivity analysis via self- and cross-atmospheric injection retrievals based on nine opacity cross-sections generated with different sets of assumptions to be representative of underlying model uncertainties.

\subsection{Opacity Database \& Cross-Sections}

We use molecular data from the HITRAN2016\cite{hitran2016}, HITEMP2010\cite{HITEMP2010} and ExoMol\cite{exomol2020} databases. The line-by-line section of the HITRAN database contains important spectroscopic parameters for individual transitions. These parameters include (but are not limited to) line position, line intensity (and Einstein-A coefficient), half-widths (and their temperature dependences) associated with the pressure broadening by different ambient gases, quantum identifications of the transition, and degeneracy factors of the energy states involved in the transition, along with the energy value of the lower state (see ref.\cite{hitran2004} for the description of some of the main parameters). The HITRAN database focuses on spectra at the terrestrial atmosphere temperatures and pressures. Therefore for many molecules, it can not be accurately used at elevated temperatures, primarily due to the paucity of transitions that become stronger when the temperature increases. The HITEMP database\cite{HITEMP2010} is similar to HITRAN in its parametrizations, however, it contains substantially more transitions for the eight molecules that are currently provided, which is referred to as the completeness issue. This allows for the modeling of spectra at elevated temperatures. The \texttt{ExoMol} database\cite{exomol2020} provides line positions, Einstein-A coefficients, and rotational quanta assignments and has thus also been addressing the completeness of line lists. The molecular data from the ExoMol database contains many orders of magnitude more transitions than those derived from experiments, due to its theoretical approach to calculating line lists. In many cases the parameters in ExoMol and HITRAN agree very well, in fact, many parameters in HITRAN are from ExoMol and vice versa. Nevertheless, in some cases, significant discrepancies between the two databases have been reported (e.g., for methane\cite{hargreaves2020}).

For atmospheric retrieval, the real-time generation of opacity cross-sections at present is computationally infeasible. Therefore, the field standard is to pre-compute the cross-sections. The generation of an opacity cross-section requires a series of assumptions. We, therefore, generated several cross-sections representing such assumptions to explore their impacts on the retrieved parameters. For this study, we generated nine equivalent cross-sections for seven different molecules. Supplementary Table \ref{tab:CS_parameter} introduces the parameters used. The cross-sections are generated using TierraCrossSection, which re-purposes functions from HAPI\cite{kochanov2016} to enable the generation of cross-sections under a wider range of assumptions. Such cross-sections are relevant for sensitivity analysis similar to the one introduced here. To generate cross-sections from the ExoMol database,  we use \texttt{exocross}\cite{exocross} for producing files with line intensities and line positions, which can be used by \texttt{TierraCrossSection}. We introduce below the nine different cross-sections and their purpose in our sensitivity analysis.

\begin{enumerate}
    \item \texttt{CS-DFLT}\textbf{:} This cross-section represents the standard for our study and is based on the default calculation options of \texttt{hapi} (v 1.0).  We use the Voigt profile for lines, air broadening parameters for pressure broadening, and calculate the extent of the line profile to 50 Voigt half widths at half maximum (HWHM).  

    \item\texttt{CS-1SUP}\textbf{:} This cross-section aims at assessing the effect of uncertainties on line parameters on retrieval capabilities. We thus modify line intensity, pressure broadening, line position, and temperature dependence of air broadening by +1$\sigma$ error. The uncertainties for the aforementioned parameters are reported in HITRAN. When the values are not available (i.e., corresponding to uncertainty index 0), we assume an error of 20 percent, although in some cases it maybe too conservative as, for instance, some of the excellent ozone data in HITRAN2016 is from the times where uncertainties were not provided in the database.  Supplementary Figure \ref{fig:lineparamuncertainty} shows the distribution of uncertainties on the line parameters for four different molecules. It highlights the wide range of uncertainties each parameters span and their correlation to the line intensity. Using these cross-sections, we can understand the impact of the current accuracy of these lines.  
    
    \item \texttt{CS-1SDN}\textbf{:} Similarly to \texttt{CS-1SUP}, the \texttt{CS-1SDN} cross-section aims at assessing the effect of uncertainties on line parameters on retrieval capabilities. Here the aforementioned line parameters are perturbed by -1$\sigma$.
    
    \item\texttt{CS-SELF}\textbf{:} This cross-section aims at assessing the effects of (1) unknown background broadeners and (2) unknown broadening parameters for the dominant broadener. To that end, instead of using air as the perturber (broadener), we use self-broadening parameters. We note that the processes behind line broadening are complex and lead to broadening parameters that are themselves a function of the rotational and vibrational quanta of the transition represented (i.e., non-monotonic dependence to wavelength). The difference between air- and self-broadening is molecule dependent, for example, some molecules such as nitrogen have a similar range of broadening coefficients, whereas for water, the broadening coefficients (self and air) range between values spanning over 0.5 dex (from 0.15 to 0.55 and from 0 to 0.15, respectively).
    
    \item \texttt{CS-500W}\textbf{:} This cross-section aims to assess the effect of imperfections in line profiles on retrieval capabilities. We generate \texttt{CS-500W} by modeling the line wings up to 500 HWHM otherwise truncated to the default value of 50. The far-wing behavior is affected by the wind patterns and turbulence in the atmosphere, phenomena not considered in current retrievals. Yet, the far wings for the strong lines can determine the overall spectral bands and opacity continuum, particularly at higher pressures.
    
    \item \texttt{CS-MAXB}\textbf{:} This cross-section aims to further explore how the pressure broadening factor can impact retrieval. We take the broadening parameter as double of the larger value between the air and the self-broadening parameter for each transition (i.e., +0.3 dex perturbation). This cross-section represents the case when the pressure broadening parameters are over-estimated. Higher pressure broadening does not change the overall opacity of the molecule, but simply redistributes the opacity from the line center into the wings. 
    
    \item \texttt{CS-MINB}\textbf{:} Similarly to \texttt{CS-MAXB}, the \texttt{CS-MINB} cross-section aims to further explore how the pressure broadening factor can impact retrieval. Here, we use half of the smaller value of each line of the broadening parameters between self-broadening and the air-broadening (i.e., -0.3 dex perturbation). This cross-section aims to reveal how underestimating the pressure broadening parameter can impact the retrieval. 
    
    \item \texttt{CS-EXML}\textbf{:} This cross-section aims at assessing the effect of differences in opacity databases used to perform a retrieval. We generate this cross-section using the ExoMol line list. To ensure the same level of completeness, we use the intensity cut-off used in HITRAN. As for the pressure broadening, we assumed a universal value of 0.07 cm$^{-1}$ with the temperature dependence exponent of 0.5 across all the molecules. ExoMol suggests a more accurate $J$-dependence for a few of the molecules, which however was not implemented to ensure the same treatment across the molecules. Molecules such as nitrogen and ozone, for which no equivalent cross-section could be created, were replaced with their equivalents from \texttt{CS-DFLT}. 
    
    \item \texttt{CS-HTMP}\textbf{:} This cross-section aims at further exploring the effect of (in)completeness of line lists comparing HITEMP2010 to HITRAN2016 as it extends the HITRAN line lists to higher temperature \cite{HITEMP2010} primarily to address the completeness issue that arises from using cutoff for the weaker lines. Molecules such as methane, and water, significant contribution towards opacity come from weaker lines at these higher temperatures. We note that both HITEMP2010 and the HITRAN2016 data are regularly updated when better experiments or theoretical calculations are performed. Hydrogen, nitrogen and ozone whose line lists are not part of HITEMP2010, their cross-sections were taken from \texttt{CS-DFLT}.
    
\end{enumerate}

\subsection{Additional Opacity}
We added two attenuation processes to our opacity models, Rayleigh scattering and collision-induced absorption (CIA). We accounted for Rayleigh scattering for all molecules except ozone (refs.\cite{dalgarno1965, sneep2005}), for which such data is unavailable and which does not occur at concentrations high enough to impact Rayleigh scattering. We accounted for CIA using the data from HITRAN CIA\cite{karman2019} for $\rm H_2-H_2$\cite{abel2011}, $\rm H_2-He$\cite{abel2012}, and $\rm N_2-N_2$\cite{lafferty1996, karman2015, hartmann2017}.

\subsection{Molecule Selection}
Interpreting a transmission spectrum requires an inventory of molecular opacities, often complex molecular networks that contribute to a final spectrum\cite{sousa2019}. For this exploratory work, we consider seven molecules that have traditionally been important in the context of planetary and exoplanetary sciences: water, methane, carbon dioxide, carbon monoxide, ozone, hydrogen, and nitrogen. Our work with these seven molecules should be representative of the typical systematic problems with any other molecules, and represent the prominent molecules for hydrogen, carbon, and oxygen chemistry. Among the seven molecules, hydrogen (H$_2$) and nitrogen (N$_2$) do not have strong spectroscopic features. Similarly, we add helium as a background gas at a constant volume mixing ratio of 15 to 85 with respect to  hydrogen\cite{conrath1987, fouchet2003}.

\subsection{Spectral Resolution}

It is necessary to simulate the planetary spectrum at a spectral resolution that is higher than the one offered by the observing instrument and then bin down to a lower resolution. Observations at R$\sim$10 with \textit{Hubble} require model evaluation at R$\sim$1,000 for example\cite{zhang2019}. For our applications focusing on \textit{JWST} observations, we tested the retrieval performances using resolutions of 10,000, 100,000, and 2,000,000. We find that the R$\sim$10,000 model leads to deviations in the simulated spectra of up to 50 p.p.m. (part per million) when compared to the R$\sim$2,000,000 model, whereas for R$\sim$100,000, the mean difference is below 10 p.p.m. Most significantly, we find that cross-retrievals using the two later resolutions lead to consistent results (while using R$\sim$10,000 lead to deviations above the 3$\sigma$ level), implying equivalent performance for R$\sim$100,000 and R$\sim$2,000,000 for the intended application. We thus choose to use R$\sim$100,000 to mitigate the contribution of the spectral resolution choice to the uncertainty budget while keeping our models computationally effective, as in ref.\cite{dewit2013}.

We note that there is an intrinsic maximum resolution attainable with a given line list due to the uncertainty on the line position (see Supplementary Figure\,\ref{fig:lineparamuncertainty}). The line lists used here report the position of most dominant lines with uncertainties below or within the 0.01 to 0.001 per cm range. This translates into a resolution between 70,000 and 1.6 million over the wavelength range covered (0.6 to 15 microns), which is comparable to the resolution selected above based on an instrumental argument (data quality and detector resolution). This means that the precision on the line positions in current opacity line lists will not affect upcoming atmospheric retrieval with JWST.

\subsection{Grid Optimization}
\label{sec:grid}
We choose our temperature-pressure grids to minimize interpolation errors. Line profiles are more sensitive to the changes in pressure at a high-pressure regime than at a low-pressure regime. Conversely, they are more sensitive to small changes in temperature at a lower temperature compared to the higher temperature regime. We minimize the effect of interpolation error by using a non-homogeneous p-T grid (see Supplementary Table\,\ref{tab:CS_parameter}), similarly to ref.\cite{dewit2013}.

\subsection{Transmission Spectroscopy Code and Benchmarking}
\label{sec:tierra}
We develop a 1D code for transmission spectra, \texttt{tierra}  (TransmIssion spEctRoscopy of tRansiting plAnets), following the code used in ref\cite{dewit2013}. We benchmark \texttt{tierra} against \texttt{petitRadtrans}\cite{molliere2019} for different atmospheres. \texttt{petitRadtrans} has itself been benchmarked against other codes\cite{tremblin2015, baudino2015} using the recommended cross-section. We find that the two transmission spectra are in excellent agreement  (median absolute deviation of 1.9 p.p.m., $<$0.005 scale height) as shown in Supplementary Figure\,\ref{fig:BenchmarkTest}. This difference ultimately found with our sets of models is thus marginal in comparison to the effects intended to be studied here (see Figure\,\ref{fig:ModelComparison}) implying that our transmission code implementation will not contribute significantly to the uncertainty budget.

\subsection{Planet, Star, and Instrument Models}
\label{sec:NoiseModel}
For this exploratory work, we use simple models for the planet and for the star to focus on the limitation stemming from current opacity models. We assume the star does not present heterogeneities which could induce contamination at the level of the planetary spectrum\cite{rackham2018}. We assume that the planetary atmosphere can be modeled via a 1D atmosphere with elemental abundances that are independent of the altitude and a temperature profile following $T(z) = T_\infty + (T_0 - T_\infty) e^{-H_T/z}  $, where $H_T$ is the temperature scale height. We note that such models would be too simplistic to perform a robust atmospheric retrieval of JWST data\cite{rackham2018,caldas2019,macdonald2020}. Yet the related limitations are not present in our approach using injection-(cross)retrieval analysis, which uses the same planet and star models to generate the synthetic spectra and analyze them.

For the instrument model, we use a model as realistic as possible to best compare the contributions of the opacity model to the total noise budget. We, therefore, use \texttt{pandexo}\cite{pandexo} for our instrument and noise models. We consider two observational cases: (1) an Earth-sized planet around a Hmag~10 M-dwarf star and (2) a Jupiter-sized planet around a late Hmag~10 K-dwarf star (see Supplementary Table\,\ref{tab:PlanetParams}). We assume in both cases combined observations with Near Infrared Spectrograph (NIRSpec) and Mid-Infrared Instrument (MIRI). In order to isolate the effect of opacity models, we assume a total of 100 transits both in  NIRSpec and MIRI with a noise floor at 10 p.p.m.\cite{schlawin2020}. We chose this large number of transits to ensure that if there is an accuracy wall relevant in the context of the JWST and ELTs era it will be revealed. And in fact, we find that for the synthetic systems chosen around H-Mag 10 stars the accuracy wall is hit within 10 to 20 transits (see Figure~4), meaning that systems such as HD~189733~b with a H-Mag~5.6 host only requires a few transits.

\subsection{Analysis Technique}

We study the effect of perturbations to our opacity models on two data products: (1) the transmission spectrum and (2) the retrieved atmospheric parameters. We quantify the deviations between transmission spectra via the ``Spectral Statistical Distance'' (SSD) defined as $\sum \frac{ ( \rm{Base~ Model}_{\lambda} - Model_{\lambda})^2}{\sigma_\lambda^2}$, where Base model is the synthetic spectrum calculated using the default cross-section (i.e., \texttt{CS-DFLT}), Model is the synthetic spectrum generated using another cross-section based on the same atmospheric parameters, and $\sigma_{\lambda}$ is the noise model associated with these observation. The ``Spectral Statistical Distance'' corresponds to the Deviation Quotient used as detection metrics in previous studies\cite{sousa2020, lincowski2018}. We later show how these metrics come with limitations.  

In order to quantify the deviation between the retrieved parameters we need a framework to derive their best estimates and uncertainties from a given spectrum and introduce the ``Parametric Statistical Distance''. We use Monte Carlo Markov Chain (MCMC) technique to assess the Posterior Probability Distributions (PPDs) of the atmospheric parameters associated with a spectrum. For these retrievals, we fixed planetary mass and radius as well as stellar ones to focus on the effect of opacity models. Our jump parameter set consists of three parameters for the temperature profile and seven molecular number densities taken at the reference radius following the formulation introduced in ref\cite{dewit2013}.

To expedite the fitting process and minimize the burn-in, we initiate the MCMCs using the original values used to generate the synthetic model with \texttt{CS-DFLT}. We use affine invariant MCMC as implemented in \texttt{emcee}\cite{emcee} for exploring the parameter space, and ran our models with 20 walkers for a minimum of 25,000 steps. Our walkers are in log-space for the number densities and the temperature lapse rate. We compute the base pressure from the number densities of the molecules and allow it to vary up to 100 atmospheres. We ensured the convergence of the MCMCs by checking for the good mixing of the walkers, and the evolution of the log probability.

Using the PPDs we can derive a ``Parametric Statistical Distance'' (PSD) to quantify the deviation in the retrieved parameters introduced by the perturbations in the opacity models. To do so, we used the Bhattacharya distance\cite{bhattacharya1943}. By measuring PSD, we can quantify biases between PPDs in a multi-dimensional fashion (see Figure~4 and Supplementary Figure \ref{fig:SE_CornerPlot}, and \ref{fig:HJ_CornerPlot}). The metric increases proportionally as the mean deviates from one another and accounts for the covariance among the parameters. If the posterior distributions are identical, the distance between them would be zero. 

We highlight the need for assessing model-driven accuracy limit at the level of final data products (here, retrieved atmospheric parameters) rather than intermediate data product (here, transmission spectrum) via the non-monotonic relationship between PSD and SSD. While SSD appears to correlate positively with PSD in log-log space, the large spread in value seen around their mean relationship highlights that perturbations at the level of a transmission spectrum cannot readily be translated into perturbations at the level of atmospheric parameters. Cross-sections initially yielding a large SSD may still enable a good fit (reduced $\chi^2\sim1$) owing to efficient compensation mechanisms. For example in the super-Earth scenario, while \texttt{CS-500W} and \texttt{CS-MAXB} have SSDs over three times those of \texttt{CS-1SUP} and \texttt{CS-1SDN} they similarly yield reduced $\chi^2$ around 1 owing to larger compensations (ie., biases) on the atmospheric parameters resulting in larger PSDs (Supplementary Table 3).

\subsection{Super-Earth Retrievals}
In using \texttt{CS-1SUP}, we find that most values are retrieved within less than 2.5$\sigma$, except for carbon dioxide which deviate by 3.18$\sigma$ (see Supplementary Table\,\ref{tab:RetrievedParams_SE}). For \texttt{CS-1SDN}, we find four parameters: $T_0$, along with molecular abundance of methane, ozone and hydrogen biased by 2.87$\sigma$,  4.02$\sigma$, 2.87$\sigma$ and -4.51$\sigma$ respectively. While no significant precision change was seen for \texttt{CS-1SUP}, retrievals from \texttt{CS-1SDN} lead to a significant precision change for $T_0$, lapse rate, and $T_\infty$ by a factor of 0.35, 2.47, and 14.49 respectively. \texttt{CS-SELF} retrievals show strong biases among temperature parameters $T_0$ (-4.02$\sigma$), and similarly are the abundances of water (3.31$\sigma$), carbon dioxide (6.91$\sigma$), methane (7.14$\sigma$), and ozone (6.11$\sigma$).  All three temperature parameters exhibit strong biases for \texttt{CS-500W}: $T_0$ (-6.62$\sigma$),  lapse rate (2.67$\sigma$), and $T_\infty$ (5.48$\sigma$). Similarly, molecular abundances of carbon monoxide (-3.23$\sigma$), water (8.09$\sigma$), carbon dioxide (8.08$\sigma$), methane (10.11$\sigma$), and ozone (9.83$\sigma$). Such biases are accompanied by greater than 2$\times$ change in the precision for lapse rate (0.34), $T_\infty$ (0.44), and  the abundances of nitrogen (39.13) and hydrogen (73.55). Retrievals using \texttt{CS-MAXB} show significant biases in $T_0$ (-6.61$\sigma$), lapse rate (3.0$\sigma$), molecular abundance of water (6.13$\sigma$), carbon dioxide (9.19$\sigma$), methane (11.26$\sigma$), and ozone (10.23$\sigma$) as shown in Supplementary Figure\,\ref{fig:SE_CornerPlot}, while retrievals using \texttt{CS-MINB} show significant biases in $T_0$ (6.19$\sigma$), and molecular abundance of ozone (-4.41$\sigma$). \texttt{CS-MAXB} led to change in the precision of posterior distribution for abundances of lapse rate (0.38), nitrogen (2.89), and hydrogen (2.9), and similarly for \texttt{CS-MINB} led to change in precision for $T_0$ (0.29), lapse rate (2.78), T$_\infty$ (15.61), and molecular abundances of hydrogen (0.45). \texttt{CS-EXML} shows biases greater than 2$\sigma$ for all of the parameters with the largest bias seen for the retrieved abundance of the methane (-17.99$\sigma$): $T_0$ (-18.18$\sigma$, and precision factor of 0.31), lapse rate (precision factor of 4.66), $T_\infty$ (-19.37$\sigma$, and precision factor of 0.18), molecular abundance of nitrogen (precision factor of 0.21), carbon monoxide (2.73$\sigma$), water (-8.32$\sigma$), carbon dioxide (-12.84$\sigma$), ozone (-6.61$\sigma$) and hydrogen (precision factor 0.4). Similarly to \texttt{CS-EXML}, \texttt{CS-HTMP} shows the strongest biases, due to a difference in completeness with \texttt{CS-DFLT}. We find that the cross-retrieval with \texttt{CS-HTMP} leads to PPDs often narrower than those of \texttt{CS-EXML} as well as an atypical PPD for CO (see last subsection for discussion).

\subsection{Warm-Jupiter Retrievals}
Using \texttt{CS-1SUP}, most parameters are retrieved within less 2.5$\sigma$, except for significant lapse rate (a factor of 2.4), see Supplementary Table\,\ref{tab:RetrievedParams_HJ}. For \texttt{CS-1SDN}, we similarly observe lapse rate a significant precision change of lapse rate (2.32). Similar to the case of super-Earth, \texttt{CS-SELF} comparatively exhibit stronger biases among numerous parameters as shown in Supplementary Figure\,\ref{fig:HJ_CornerPlot}: $T_0$ (-7.54$\sigma$), lapse rate (2.87$\sigma$), molecular abundances of carbon dioxide (3.37$\sigma$), and ozone (2.54$\sigma$).  \texttt{CS-500W} retrievals exhibited strong biases for $T_0$ (-7.89$\sigma$), lapse rate (precision change factor of 2.12), and abundances of carbon monoxide (-5.77$\sigma$). Retrievals using \texttt{CS-MAXB} showed significant biases in $T_0$ (-25.68$\sigma$), lapse rate (3.85$\sigma$), molecular abundance of carbon monoxide (3.73$\sigma$), and ozone (2.73$\sigma$), while retrievals using \texttt{CS-MINB} show significant biases just for molecular abundance of methane (-2.60$\sigma$) and hydrogen (-11.1$\sigma$). The change in precision when using \texttt{CS-MAXB} is for $T_0$ (0.2) whereas for \texttt{CS-MINB} it is observed for lapse rate (11.85), $T_\infty$ (2.54). and hydrogen (0.25). Using \texttt{CS-EXML} yield biases for the warm-Jupiter for methane (-11.2$\sigma$), and significant change in the precision for lapse rate (10.85), and $T_{\infty} (11.36)$. As for the super-Earth case, \texttt{CS-HTMP} and \texttt{CS-EXML} similarly show the strongest biases, due to a primary difference \texttt{CS-DFLT} in the form of increased completeness.

\subsection{Information Content and Bias Propagation in Transmission Spectra}
A planetary transmission spectrum encodes information on the pressure, temperature, and the number densities at the pressure levels probed, which thus also includes their dependence with altitude via quantities such as the pressure scale height\cite{dewit2013}. This complex information encoding is guided by the opacity model, resulting in a complex $\chi^2$ topology that is strongly sensitive to the assumptions and uncertainties of opacity models (see e.g., Figure~3 and Supplementary Figure\,\ref{fig:HJParameters}). In this subsection, we focus on two specific aspects of our sensitivity analysis to provide the reader with an increased understanding of the opacity challenge. First, we look at the uniquely sharp PPD of CO for the \texttt{CS-HTMP} retrieval of the super-Earth case (see Figure~3, top right panel). Then, we turn to the correlation between atmospheric parameters highlighted notably in Supplementary Figure\,\ref{fig:SE_CornerPlot}.

The top right panel of Figure~3 highlights that \texttt{CS-HTMP} yields a uniquely tight constraint on the base-pressure number density of CO (Log$_{10}$N$_{\rm CO}$). We leverage this unique characteristic to provide insights into why \texttt{CS-HTMP} yields a topology of the $\chi^2$ distribution for the super-Earth case resulting in such a different constraint on CO. For the case of the super-Earth, \texttt{CS-HTMP} and \texttt{CS-EXML} lead to the strongest shift (and bias) towards lower temperatures and lower number densities for all molecules but N$_2$, H$_2$, and CO. This is due to the increased level of completeness of these cross-sections yielding higher opacity in the valleys between molecular features and resulting in biases in perceived number densities and base pressure. Indeed, the same level of absorption in these valleys requires an overall lower base pressure and number density for the related absorber. We note that this effect dominates the perturbations induced in the cross-retrievals with \texttt{CS-HTMP} and \texttt{CS-EXML} for the super-Earth case but not for the warm-Jupiter case owing to the absence of scattering and CIA for the former, which exacerbates its sensitivity to deeper inter-feature valleys in the VIS-NIR region of the spectra (see Figure~3 vs Supplementary Figure\,\ref{fig:CO_Perturbation}). Second, \texttt{CS-HTMP} yields smaller number densities than \texttt{CS-EXML} for all but H$_2$ and CO, resulting in CO bearing more of the constraints associated to the atmospheric scale height (via the mean molecular weight) and the base pressure. Supplementary Figure\,\ref{fig:CO_Perturbation} highlights this aspect by presenting the effect of perturbing Log$_{10}$N$_{\rm CO}$ from the best-fit value for both \texttt{CS-DFLT} and \texttt{CS-HTMP} by -0.25, which correspond to the 1$\sigma$ confidence interval for Log$_{10}$N$_{\rm CO}$ with \texttt{CS-DFLT} (Supplementary Table\,\ref{tab:CS_parameter}). We show in the lower panel that doing so results in a worsening of the $\chi^2$ over three and a half times more significant for \texttt{CS-HTMP} than for \texttt{CS-DFLT}, echoing the differences seen in the top right panel of Figure~3 (namely a tighter constraint on CO). In addition, the bottom panel of Supplementary Figure\,\ref{fig:CO_Perturbation} highlights that in both cases the primary effect of perturbing Log$_{10}$N$_{\rm CO}$ for this super-Earth case leads to amplifying/shrinking all molecular features more so than only changing the molecular features of CO, meaning that the primary effect of such a perturbation is indeed on the mean molecular weight (and to a lesser extent the base pressure).

In Supplementary Figure\,\ref{fig:SE_CornerPlot}, we report the PPDs of the retrieved parameters for the case of the super-Earth for self-retrieval (\texttt{CS-DFLT}, red) and cross-retrieval (\texttt{CS-MAXB}, green). The only difference between the two cross-sections relates to broadening; \texttt{CS-DFLT} assumes air-broadening (geocentric) while  \texttt{CS-MAXB} assumes twice the maximum between air- and self-broadening. This perturbation is only impacting pressure levels at which pressure-related broadening effect dominate temperature-related broadening effect. Therefore, above $\sim$0.1 mbar the p-T profiles retrieved are consistent, and $T_\infty$ is not biased (see top right panel). At deeper levels, the increased broadening for \texttt{CS-MAXB} implies--all other things being constant--a decrease of line intensities and an increase in line width. The primary compensation mechanism for this perturbation is an increase in the number densities for molecules with strong molecular features and a decrease in temperature. This explains the trend seen between $T_0$ and CO$_2$, CH$_4$, H$_2$O and O$_3$ within and between the PPDs of \texttt{CS-DFLT} and \texttt{CS-MAXB}.

\newpage
\pagebreak
\clearpage

\section{Captions of Figures}

\textbf{Figure 1. Ensemble of opacity-model perturbations at the level of an opacity cross-section.} \textbf{a:} Calculated cross-sections showing the 1.3663 micron water line at 450 K, 0.1 atm, and 2-million resolution. Each cross-section is identified by its color and label on the right and is generated following specific assumptions on parameters such as broadening coefficients, far-wing behavior, and data source (see section Opacity Database \& Cross-Sections in Methods). The 1$\sigma$ envelope is the bound between CS-1SUP and CS-1SDN, which respectively account for the +1 and -1$\sigma$ perturbations on line parameters stemming from lab measurements (incl., pressure broadening and line intensity and position). \texttt{CS-EXML} based on ExoMol\cite{exomol2020} differs from the other models at pressure above $\sim$1~mbar due to the absence of pressure shift on the line positions. \textbf{b:} Calculated cross-sections showing 1.4 micron water feature at 450 K and 100 atm.

\textbf{Figure 2. Propagation of the ensemble of opacity-model perturbations to the level of a transmission spectrum.} Comparison of synthetic warm-Jupiter transmission spectra for a combined broadband NIRSpec and MIRI observation for the nine different opacity models used in this perturbation/sensitivity analysis (see section Opacity Database \& Cross-Sections in Methods) (Top). Non-random differences with amplitudes ranging from 20 to 150 part per millions (p.p.m.) are observed compared to the nominal (i.e., unperturbed) \texttt{CS-DFLT} model (Bottom), which is substantial in the era of \textit{JWST}.

\textbf{Figure 3. Transmission-spectrum fit quality unaffected by opacity-model perturbations.} Best-fit model for the cross-retrieval of a warm-Jupiter from its synthetic transmission spectrum simulated with nominal cross-section \texttt{CS-DFLT} and retrieved with cross-section \texttt{CS-SELF} (see Methods) (Top). Spectroscopic features/bands caused by absorption have been indicated. Despite the significant differences among the opacity models (Figures~1 \&~2), the absence of structure in the residuals with $\chi^2_{\nu}$=1.068 (Bottom) implies that a good fit is enabled by compensation errors on the retrieved atmospheric parameters (i.e., biases). This implies that the limitations of current opacity models may remain undetected in upcoming atmospheric retrieval efforts.

\textbf{Figure 4. Propagation of the ensemble of opacity-model perturbations to the level of retrieved atmospheric properties.} Posterior probability distributions (PPDs) of the retrieved atmospheric parameters (i.e., final data product) for the super-Earth scenario highlighting the biases induced by perturbations to our opacity model. Each cross-section is identified by its color and label on the right. The dotted black vertical lines represent the true values used in generating the synthetic spectrum. Deviations with a statistical significance of up to $\sim$20$\sigma$ and physical significance of over 1~dex are reported.
This suggests the significant sensitivity of retrieved atmospheric properties to opacity models in upcoming atmospheric retrieval efforts.

\textbf{Figure 5. The 0.5-1.0 dex opacity-driven accuracy wall for exoplanet atmospheric characterization.} \textbf{a:} Relationship between the Spectral Statistical Distance (SSD) and the Parametric Statistical Distance (PSD) relating the statistical deviation between the transmission spectrum generated with a cross-section and the one generated with the nominal cross-section to the statistical deviation between the PPDs retrieved through cross- and self-retrieval. \textbf{b:} Relationship between the SSD and the Parametric Physical Distance revealing biases between 0.5 to 1 dex on the retrieved parameters, labelled ``Opacity-driven Accuracy Wall''. This translates into an accuracy drop from what the instrument alone could deliver as highlighted by the shaded green area for our observation scenario and the dotted line for Cycle 1 Program.  The filled markers show the maximum distance amongst the retrieved atmospheric parameters, the open markers their median distance.

\begin{addendum}
 \item P.N. acknowledges the support of the Grayce B. Kerr Fellowship Fund and Elliot Fellowship at MIT. The authors would like to thank Dr. B. V. Rackham, Dr. E. K. Conway, and Prof. S. Seager for discussions on various topics presented in this paper. P. N. would like to thank Dr. P. Molliere for benchmark testing with \texttt{petitRadtrans}. The authors acknowledge the MIT SuperCloud and Lincoln Laboratory Supercomputing Center for providing (HPC, database, consultation) resources that have contributed to the research results reported within this paper. The authors express gratitude to the reviewers for their comments and suggestions that resulted in a substantial improvement of the manuscript.
 
\item[Author Contributions] J.d.W. designed the study. P.N. developed the computational framework for the study with the support of J.d.W., I.E.G., R.J.H., C.S.-S., and R.V.K., P.N. and J.d.W. led the analysis and interpretation with the support of I.E.G., R.J.H., and C.S.-S. I.E.G., R.J.H. and C.S.-S. lead the discussions regarding future avenues for improving opacity models. Every author contributed to writing the manuscript. 

\item[Competing Interests] The authors declare that they have no competing financial interests.

\item[Data Availability]
This paper makes use of the opacity data from the HITRAN2016\cite{hitran2016}, HITEMP2010\cite{HITEMP2010} and ExoMol\cite{exomol2020} databases. The perturbed opacity cross-sections and the synthetic exoplanet transmission spectra are available from the corresponding author upon request and can be (re)generated using the \texttt{tierraCrossSection} and \texttt{tierra} codes (see Code Availability), respectively.

\item[Code Availability] This work makes use of the following publicly available codes: \texttt{emcee}\cite{emcee},  \texttt{ExoCross}\cite{exocross}, \texttt{hapi}\cite{kochanov2016}.  Additionally, it is based on \texttt{tierra} which we is now publicly available at {https://github.com/disruptiveplanets /tierra}, and \texttt{tierraCrossSection} which is also available at {https://github.com/disruptiveplanets/tierraCrossSection}. 

 \item[Correspondence] Correspondence and requests for materials
should be addressed to JdW~(email: jdewit@mit.edu). 

\end{addendum}

\section*{References}

\begin{thebibliography}{100}
\expandafter\ifx\csname url\endcsname\relax
  \def\url#1{\texttt{#1}}\fi
\expandafter\ifx\csname urlprefix\endcsname\relax\def\urlprefix{URL }\fi
\providecommand{\bibinfo}[2]{#2}
\providecommand{\eprint}[2][]{\url{#2}}

\makeatletter
\makeatother

\bibitem{rackham2018}
\bibinfo{author}{Rackham, B.V., Apai, D. \& Giampapa, M.S.}
\newblock \bibinfo{title}{{The Transit Light Source Effect: False Spectral Features and Incorrect Densities for M-dwarf Transiting Planets}}.
\newblock \emph{\bibinfo{journal}{\apj}}
  \textbf{\bibinfo{volume}{853}}, \bibinfo{pages}{122} (\bibinfo{year}{2018}).

\bibitem{macdonald2020}
\bibinfo{author}{MacDonald, R.J., Goyal, J.M. \& Lewis, N.K.}
\newblock \bibinfo{title}{{Why Is it So Cold in Here? Explaining the Cold Temperatures Retrieved from Transmission Spectra of Exoplanet Atmospheres}}.
\newblock \emph{\bibinfo{journal}{\apj}}
  \textbf{\bibinfo{volume}{893}}, \bibinfo{pages}{L43} (\bibinfo{year}{2020}).

\bibitem{burrows2014}
\bibinfo{author}{Burrows, A.S.}
\newblock \bibinfo{title}{{Highlights in the study of exoplanet atmospheres}}.
\newblock \emph{\bibinfo{journal}{\nat}}
  \textbf{\bibinfo{volume}{513}}, \bibinfo{pages}{345-352} (\bibinfo{year}{2014}).

\bibitem{caldas2019}
\bibinfo{author}{Caldas, A., Leconte, J., Selsis, F., Waldmann, I.P., Bord{\'e}, P., Rocchetto, M., \& Charnay, B.}
\newblock \bibinfo{title}{{Effects of a fully 3D atmospheric structure on exoplanet transmission spectra: retrieval biases due to day-night temperature gradients}}.
\newblock \emph{\bibinfo{journal}{\aap}}
  \textbf{\bibinfo{volume}{623}}, \bibinfo{pages}{A161} (\bibinfo{year}{2019}).

\bibitem{betremieux2017}
\bibinfo{author}{B{\'e}tr{\'e}mieux, Y. \& Swain, M.R.}
\newblock \bibinfo{title}{{An analytical formalism accounting for clouds and other `surfaces' for exoplanet transmission spectroscopy}}.
\newblock \emph{\bibinfo{journal}{\mnras}}
  \textbf{\bibinfo{volume}{467}}, \bibinfo{pages}{2834–2844} (\bibinfo{year}{2017}).

\bibitem{barstow2020}
\bibinfo{author}{Barstow, J.K.}
\newblock \bibinfo{title}{{Unveiling cloudy exoplanets: the influence of cloud model choices on retrieval solutions}}.
\newblock \emph{\bibinfo{journal}{\mnras}}
  \textbf{\bibinfo{volume}{497}}, \bibinfo{pages}{4183-4195} (\bibinfo{year}{2020}).

\bibitem{hu2012}
\bibinfo{author}{Hu, R., Seager, S. \& Bains, W.}
\newblock \bibinfo{title}{{Photochemistry in Terrestrial Exoplanet Atmospheres. I. Photochemistry Model and Benchmark Cases}}.
\newblock \emph{\bibinfo{journal}{\apj}}
  \textbf{\bibinfo{volume}{761}}, \bibinfo{pages}{166} (\bibinfo{year}{2012}).

\bibitem{heng2017}
\bibinfo{author}{Heng, K., \& Kitzmann, D.}
\newblock \bibinfo{title}{{The theory of transmission spectra revisited: a semi-analytical method for interpreting WFC3 data and an unresolved challenge}}.
\newblock \emph{\bibinfo{journal}{\mnras}}
  \textbf{\bibinfo{volume}{470}}, \bibinfo{pages}{2972-2981} (\bibinfo{year}{2017}).
  
  
\bibitem{welbanks2019}
\bibinfo{author}{Welbanks, L., \& Madhusudhan, N.}
\newblock \bibinfo{title}{{On Degeneracies in Retrievals of Exoplanetary Transmission Spectra}}.
\newblock \emph{\bibinfo{journal}{\aj}}
  \textbf{\bibinfo{volume}{157}}, \bibinfo{pages}{206} (\bibinfo{year}{2019}).


\bibitem{dePater2005}
\bibinfo{author}{de Pater, I., DeBoer, D., Marley, M., Freedman, R. \& Young, R.}
\newblock \bibinfo{title}{{Retrieval of water in Jupiter's deep atmosphere using microwave spectra of its brightness temperature}}.
\newblock \emph{\bibinfo{journal}{\icarus}}
  \textbf{\bibinfo{volume}{173}}, \bibinfo{pages}{425-438} (\bibinfo{year}{2005}).

\bibitem{hedges2016}
\bibinfo{author}{Hedges, C. \& Madhusudhan, N.}
\newblock \bibinfo{title}{{Effect of pressure broadening on molecular absorption cross sections in exoplanetary atmospheres}}.
\newblock \emph{\bibinfo{journal}{\mnras}}
  \textbf{\bibinfo{volume}{458}}, \bibinfo{pages}{1427-1449} (\bibinfo{year}{2016}).


\bibitem{baudino2017}
\bibinfo{author}{Baudino, J.-L.} \emph{et~al.}
\newblock \bibinfo{title}{Toward the Analysis of JWST  Exoplanet Spectra: Identifying Troublesome Model Parameters}.
\newblock \emph{\bibinfo{journal}{\apj}}
  \textbf{\bibinfo{volume}{850}}, \bibinfo{pages}{150} (\bibinfo{year}{2017}).


\bibitem{nezhad2019}
\bibinfo{author}{Gharib-Nezhad, E. \& Line, M.R.}
\newblock \bibinfo{title}{The Influence of H$_{2}$O Pressure Broadening in High-metallicity Exoplanet Atmospheres}.
\newblock \emph{\bibinfo{journal}{\apj}}
  \textbf{\bibinfo{volume}{872}}, \bibinfo{pages}{27} (\bibinfo{year}{2019}).


\bibitem{greaves2021}
\bibinfo{author}{Greaves, E.} \emph{et~al.}
\newblock \bibinfo{title}{Phosphine gas in the cloud decks of Venus}.
\newblock \emph{\bibinfo{journal}{Nature Astronomy}}
  \textbf{\bibinfo{volume}{5}}, \bibinfo{pages}{655-664} (\bibinfo{year}{2021}).
  
\bibitem{ranjan2020}
\bibinfo{author}{Ranjan, S.} \emph{et~al.}
\newblock \bibinfo{title}{Photochemistry of Anoxic Abiotic Habitable Planet Atmospheres: Impact of New H$_{2}$O Cross Sections}.
\newblock \emph{\bibinfo{journal}{\apj}}
  \textbf{\bibinfo{volume}{896}}, \bibinfo{pages}{148} (\bibinfo{year}{2020}).

\bibitem{fortney2019}
\bibinfo{author}{Fortney, J.} \emph{et~al.}
\newblock \bibinfo{title}{The Need for Laboratory Measurements and Ab Initio Studies to Aid Understanding of Exoplanetary Atmospheres}.
\newblock \emph{\bibinfo{journal}{\apj}}
  \textbf{\bibinfo{volume}{2020}}, \bibinfo{pages}{146} (\bibinfo{year}{2019}).


\bibitem{giacobbe2021}
\bibinfo{author}{Giacobbe, P.} \emph{et~al.}
\newblock \bibinfo{title}{Five carbon- and nitrogen-bearing species in a hot giant planet's atmosphere}.
\newblock \emph{\bibinfo{journal}{\nat}}
  \textbf{\bibinfo{volume}{592}}, \bibinfo{pages}{205-208} (\bibinfo{year}{2021}).


\bibitem{batalha2017}
\bibinfo{author}{Batalha, N. \& Line, M.R.} 
\newblock \bibinfo{title}{Information Content Analysis for Selection of Optimal JWST Observing Modes for Transiting Exoplanet Atmospheres}.
\newblock \emph{\bibinfo{journal}{\aj}}
  \textbf{\bibinfo{volume}{153}}, \bibinfo{pages}{151} (\bibinfo{year}{2017}).

\bibitem{kochanov2016}
\bibinfo{author}{Kochanov, R.V.} \emph{et~al.} 
\newblock \bibinfo{title}{HITRAN Application Programming Interface (HAPI): A comprehensive approach to working with spectroscopic data}.
\newblock \emph{\bibinfo{journal}{\jqsrt}}
  \textbf{\bibinfo{volume}{177}}, \bibinfo{pages}{15-30} (\bibinfo{year}{2016}).

\bibitem{hitran2016}
\bibinfo{author}{Gordon, I.E.} \emph{et~al.} 
\newblock \bibinfo{title}{The HITRAN2016 molecular spectroscopic database}.
\newblock \emph{\bibinfo{journal}{\jqsrt}}
  \textbf{\bibinfo{volume}{203}}, \bibinfo{pages}{3-69} (\bibinfo{year}{2017}).

\bibitem{exomol2020}
\bibinfo{author}{Tennyson, J.} \emph{et~al.} 
\newblock \bibinfo{title}{The 2020 release of the ExoMol database: Molecular line lists for exoplanet and other hot atmospheres}.
\newblock \emph{\bibinfo{journal}{\jqsrt}}
  \textbf{\bibinfo{volume}{255}}, \bibinfo{pages}{107-228} (\bibinfo{year}{2020}).
  
\bibitem{HITEMP2010}
\bibinfo{author}{Rothman, L.S.} \emph{et~al.} 
\newblock \bibinfo{title}{HITEMP, the high-temperature molecular spectroscopic database}.
\newblock \emph{\bibinfo{journal}{\jqsrt}}
  \textbf{\bibinfo{volume}{111}}, \bibinfo{pages}{2139-2150} (\bibinfo{year}{2010}).
  
\bibitem{dewit2013}
\bibinfo{author}{de Wit, J. \& Seager, S.}
\newblock \bibinfo{title}{Constraining Exoplanet Mass from Transmission Spectroscopy}.
\newblock \emph{\bibinfo{journal}{Science}}
  \textbf{\bibinfo{volume}{342}}, \bibinfo{pages}{1473-1477} (\bibinfo{year}{2013}).


\bibitem{bhattacharya1943}
\bibinfo{author}{Bhattacharya, A.}
\newblock \bibinfo{title}{On a Measure of Divergence between Two Statistical Populations Defined by Their Probability Distributions}.
\newblock \emph{\bibinfo{journal}{Bulletin of the Calcutta Mathematical Society}}
  \textbf{\bibinfo{volume}{160}}, \bibinfo{pages}{99-109} (\bibinfo{year}{1943}).

\bibitem{hargreaves2020}
\bibinfo{author}{Hargreaves, R.J.} \emph{et~al.} 
\newblock \bibinfo{title}{An Accurate, Extensive, and Practical Line List of Methane for the HITEMP Database}.
\newblock \emph{\bibinfo{journal}{\apjs}}
  \textbf{\bibinfo{volume}{247}}, \bibinfo{pages}{55} (\bibinfo{year}{2020}).
  
  \bibitem{10.1016/j.jqsrt.2015.09.003}
\bibinfo{author}{Wilzewski, J.S.} \emph{et~al.} 
\newblock \bibinfo{title}{H$_{2}$, He, and CO$_{2}$ line-broadening coefficients, pressure shifts and temperature-dependence exponents for the HITRAN database. Part 1: SO$_{2}$, NH$_{3}$, HF, HCl, OCS and C$_{2}$H$_{2}$}.
\newblock \emph{\bibinfo{journal}{\jqsrt}}
  \textbf{\bibinfo{volume}{168}}, \bibinfo{pages}{193-206} (\bibinfo{year}{2016}).

  \bibitem{10.1029/2019JD030929}
\bibinfo{author}{Tan, Y.} \emph{et~al.} 
\newblock \bibinfo{title}{Introduction of water-vapor broadening parameters and their temperature-dependent exponents into the HITRAN database: Part I{\textemdash}CO$_{2}$, N$_{2}$O, CO, CH$_{4}$, O$_{2}$, NH$_{3}$, and H$_{2}$S}.
\newblock \emph{\bibinfo{journal}{\jqsrt}}
  \textbf{\bibinfo{volume}{124}}, \bibinfo{pages}{11,580-11,594} (\bibinfo{year}{2019}).

  \bibitem{Krishnaji1963}
\bibinfo{author}{Krishnaji, C.}
\newblock \bibinfo{title}{Molecular Interaction and Linewidth of the Asymmetric Molecule SO$_{2}$. II. SO$_{2}$-CO$_{2}$ Collisions}.
\newblock \emph{\bibinfo{journal}{The Journal of Chemical Physics}}
  \textbf{\bibinfo{volume}{38}}, \bibinfo{pages}{1019} (\bibinfo{year}{1963}).

\bibitem{Ceselin2017}
\bibinfo{author}{Ceselin, G.} \emph{et~al.} 
\newblock \bibinfo{title}{CO$_2$-, He- and H$_2$-broadening coefficients of SO$_2$ for $\nu$1 band and ground state transitions for astrophysical applications}.
\newblock \emph{\bibinfo{journal}{\jqsrt}}
  \textbf{\bibinfo{volume}{203}}, \bibinfo{pages}{367-376} (\bibinfo{year}{2017}).
  
  
\bibitem{Dudaryonok2018}
\bibinfo{author}{Dudaryonok, A.S. \& Lavrentieva, N.N.} 
\newblock \bibinfo{title}{Theoretical estimation of SO$_2$ line broadening coefficients induced by carbon dioxide in the 150-300 K temperature range}.
\newblock \emph{\bibinfo{journal}{\jqsrt}}
  \textbf{\bibinfo{volume}{219}}, \bibinfo{pages}{360-365} (\bibinfo{year}{2018}).
  
\bibitem{10.1016/j.jqsrt.2018.12.030}
\bibinfo{author}{Borkov, Y.G.} \emph{et~al.} 
\newblock \bibinfo{title}{CO$_{2}$-broadening and shift coefficients of sulfur dioxide near 4 {\ensuremath{\mu}}m}.
\newblock \emph{\bibinfo{journal}{\jqsrt}}
  \textbf{\bibinfo{volume}{225}}, \bibinfo{pages}{119-124} (\bibinfo{year}{2019}).
  
\bibitem{10.1016/j.jqsrt.2020.107283}
\bibinfo{author}{Hashemi, R.} \emph{et~al.} 
\newblock \bibinfo{title}{Revising the line-shape parameters for air- and self-broadened CO$_2$ lines toward a sub-percent accuracy level}.
\newblock \emph{\bibinfo{journal}{\jqsrt}}
  \textbf{\bibinfo{volume}{256}}, \bibinfo{pages}{107283} (\bibinfo{year}{2020}).

\bibitem{Gordon2007}
\bibinfo{author}{Gordon, I.E.} \emph{et~al.} 
\newblock \bibinfo{title}{Current updates of the water-vapor line list in HITRAN: A new ``Diet'' for air-broadened half-widths}.
\newblock \emph{\bibinfo{journal}{\jqsrt}}
  \textbf{\bibinfo{volume}{108}}, \bibinfo{pages}{389--402} (\bibinfo{year}{2017}).

\bibitem{Nguyen2020}
\bibinfo{author}{Nguyen, H.T.} \emph{et~al.} 
\newblock \bibinfo{title}{Line-shape parameters and their temperature dependence predicted from molecular dynamics simulations for O$_2$- and air-broadened CO$_2$ lines}.
\newblock \emph{\bibinfo{journal}{\jqsrt}}
  \textbf{\bibinfo{volume}{242}}, \bibinfo{pages}{106729} (\bibinfo{year}{2020}).

\bibitem{Jozwiak2021}
\bibinfo{author}{J{\'{o}}{\'{z}}wiak, H.} \emph{et~al.} 
\newblock \bibinfo{title}{Line-shape parameters and their temperature dependence predicted from molecular dynamics simulations for O$_2$- and air-broadened CO$_2$ lines}.
\newblock \emph{\bibinfo{journal}{The Journal of Chemical Physics}}
  \textbf{\bibinfo{volume}{154}}, \bibinfo{pages}{054314} (\bibinfo{year}{2021}).


\bibitem{hartmann2002}
\bibinfo{author}{Hartmann, J.-M.} \emph{et~al.} 
\newblock \bibinfo{title}{A far wing lineshape for $H_2$ broadened $CH_4$ infrared transitions}.
\newblock \emph{\bibinfo{journal}{\jqsrt}}
  \textbf{\bibinfo{volume}{72}}, \bibinfo{pages}{117} (\bibinfo{year}{2002}).

\bibitem{Mlawer2012}
\bibinfo{author}{Mlawer, E.J.} \emph{et~al.} 
\newblock \bibinfo{title}{A far wing lineshape for $H_2$ broadened $CH_4$ infrared transitions}.
\newblock \emph{\bibinfo{journal}{Philosophical Transactions of the Royal Society A: Mathematical, Physical and Engineering Sciences}}
  \textbf{\bibinfo{volume}{370}}, \bibinfo{pages}{2520--2556} (\bibinfo{year}{2012}).


\bibitem{Cousin1985}
\bibinfo{author}{Cousin, C.} \emph{et~al.} 
\newblock \bibinfo{title}{Temperature dependence of the absorption in the region beyond the 43-$\mu$m band head of CO$_2$ 2: N$_2$ and O$_2$ broadening}.
\newblock \emph{\bibinfo{journal}{Applied Optics}}
  \textbf{\bibinfo{volume}{24}}, \bibinfo{pages}{3899} (\bibinfo{year}{1985}).

\bibitem{Toon2016}
\bibinfo{author}{Toon, G.C.} \emph{et~al.} 
\newblock \bibinfo{title}{HITRAN spectroscopy evaluation using solar occultation FTIR spectra}.
\newblock \emph{\bibinfo{journal}{\jqsrt}}
  \textbf{\bibinfo{volume}{182}}, \bibinfo{pages}{324--336} (\bibinfo{year}{2016}).

\bibitem{Olsen2019}
\bibinfo{author}{Olsen, K.S.} \emph{et~al.} 
\newblock \bibinfo{title}{Validation of the HITRAN 2016 and GEISA 2015 line lists using ACE-FTS solar occultation observations}.
\newblock \emph{\bibinfo{journal}{\jqsrt}}
  \textbf{\bibinfo{volume}{236}}, \bibinfo{pages}{106590} (\bibinfo{year}{2019}).







\bibitem{emcee}
\bibinfo{author}{Foreman-Mackey, D., Hogg, D. W., Lang, D. \& Goodman, J.}
\newblock \bibinfo{title}{{emcee: The MCMC Hammer}}.
\newblock \emph{\bibinfo{journal}{\pasp}}
  \textbf{\bibinfo{volume}{125}}, \bibinfo{pages}{306} (\bibinfo{year}{2017}).
  
\bibitem{exocross}
\bibinfo{author}{Yurchenko, S. N., Al-Refaie, A. F. \& Tennyson, J.}
\newblock \bibinfo{title}{{EXOCROSS: a general program for generating spectra from molecular line lists}}.
\newblock \emph{\bibinfo{journal}{\aap}}
  \textbf{\bibinfo{volume}{614}}, \bibinfo{pages}{A131} (\bibinfo{year}{2018}).  


\bibitem{hitran2004}
\bibinfo{author}{{Rothman}, L.~S.} \emph{et~al.}
\newblock \bibinfo{title}{{The HITRAN 2004 molecular spectroscopic database}}.
\newblock \emph{\bibinfo{journal}{\jqsrt}}
\textbf{\bibinfo{volume}{   }}, \bibinfo{pages}{139--204} (\bibinfo{year}{2005}).


\bibitem{dalgarno1965}
\bibinfo{author}{{Dalgarno}, A. \& {Williams}, D.~A.} 
\newblock \bibinfo{title}{{Properties of the hydrogen molecule}}.
\newblock \emph{\bibinfo{journal}{Proceedings of the Physical Society}}
  \textbf{\bibinfo{volume}{85}}, \bibinfo{pages}{685--689} (\bibinfo{year}{1965}).

   
\bibitem{sneep2005}
\bibinfo{author}{{Sneep}, M. \& {Ubachs}, W.} 
\newblock \bibinfo{title}{{Direct measurement of the Rayleigh scattering cross section in various gases}}.
\newblock \emph{\bibinfo{journal}{\jqsrt}}
\textbf{\bibinfo{volume}{92}}, \bibinfo{pages}{293--310} (\bibinfo{year}{2005}).

\bibitem{karman2019}
\bibinfo{author}{{Karman}, T.} \emph{et~al.}
\newblock \bibinfo{title}{{Update of the HITRAN collision-induced absorption section}}.
\newblock \emph{\bibinfo{journal}{\icarus}}
\textbf{\bibinfo{volume}{328}}, \bibinfo{pages}{160-175} (\bibinfo{year}{2019}).

\bibitem{abel2011}
\bibinfo{author}{Abel, M., Frommhold, L., Li, X. \& Hunt, K. L. C.}
\newblock \bibinfo{title}{{Collision-Induced Absorption by  H$_{2}$ Pairs: From Hundreds to Thousands of Kelvin}}.
\newblock \emph{\bibinfo{journal}{Journal of Physical Chemistry A}}
\textbf{\bibinfo{volume}{115}}, \bibinfo{pages}{6805--6812} (\bibinfo{year}{2011}).

\bibitem{abel2012}
\bibinfo{author}{Abel, M., Frommhold, L., Li, X. \& Hunt, K. L. C.}
\newblock \bibinfo{title}{{Infrared absorption by collisional H$_{2}$-He complexes at temperatures up to 9000 K and frequencies from 0 to 20 000 cm$^{-1}$}}.
\newblock \emph{\bibinfo{journal}{\jcp}}
\textbf{\bibinfo{volume}{136}}, \bibinfo{pages}{044319--044319} (\bibinfo{year}{2012}).


\bibitem{lafferty1996}
\bibinfo{author}{Lafferty, W. J., Solodov, A. M., Weber, A., Olson, W. B. \& Hartmann, J.-M. }
\newblock \bibinfo{title}{{Infrared collision-induced absorption by N$_2$ near 4.3 $\textmu$m for atmospheric applications: measurements and empirical modeling}}.
\newblock \emph{\bibinfo{journal}{\ao}}
\textbf{\bibinfo{volume}{35}}, \bibinfo{pages}{5911-5917} (\bibinfo{year}{1996}).

\bibitem{karman2015}
\bibinfo{author}{Karman, T., van der Avoird, A. \& Groenenboom, G. C. }
\newblock \bibinfo{title}{{Quantum mechanical calculation of the collision-induced absorption spectra of N$_{2}$-N$_{2}$ with anisotropic interactions}}.
\newblock \emph{\bibinfo{journal}{\jcp}}
\textbf{\bibinfo{volume}{142}}, \bibinfo{pages}{084305} (\bibinfo{year}{2015}).


\bibitem{hartmann2017}
\bibinfo{author}{{Hartmann}, J. M. and {Boulet}, C. and {Toon}, G.~C.}
\newblock \bibinfo{title}{{Collision-induced absorption by N$_{2}$ near 2.16 {\textmu}m: Calculations, model, and consequences for atmospheric remote sensing}}.
\newblock \emph{\bibinfo{journal}{Journal of Geophysical Research (Atmospheres)}}
\textbf{\bibinfo{volume}{122}}, \bibinfo{pages}{2419-2428} (\bibinfo{year}{2017}).


\bibitem{sousa2019}
\bibinfo{author}{Sousa-Silva, C., Petkowski, J. J. \& Seager, S.} 
\newblock \bibinfo{title}{{Molecular simulations for the spectroscopic detection of atmospheric gases}}.
\newblock \emph{\bibinfo{journal}{Physical Chemistry Chemical Physics (Incorporating Faraday Transactions)}}
\textbf{\bibinfo{volume}{21}}, \bibinfo{pages}{18970--18987} (\bibinfo{year}{2019}).


\bibitem{conrath1987}
\bibinfo{author}{Conrath, B., Gautier, D., Hanel, R., Lindal, G. \& Marten, A. T}
\newblock \bibinfo{title}{{The helium abundance of Uranus from Voyager measurements}}.
\newblock \emph{\bibinfo{journal}{\jgr}}
\textbf{\bibinfo{volume}{92}}, \bibinfo{pages}{15003--15010} (\bibinfo{year}{1987}).


\bibitem{fouchet2003}
\bibinfo{author}{Fouchet, T., Lellouch, E. \& Feuchtgruber, H.} 
\newblock \bibinfo{title}{{The hydrogen ortho-to-para ratio in the stratospheres of the giant planets}}.
\newblock \emph{\bibinfo{journal}{\icarus}}
\textbf{\bibinfo{volume}{161}}, \bibinfo{pages}{127--143} (\bibinfo{year}{2003}).


\bibitem{zhang2019}
\bibinfo{author}{Zhang, M., Chachan, Y., Kempton, E. M. R. \& Knutson, H. A. }
\newblock \bibinfo{title}{{Forward Modeling and Retrievals with PLATON, a Fast Open-source Tool}}.
\newblock \emph{\bibinfo{journal}{\pasp}}
 \textbf{\bibinfo{volume}{131}}, \bibinfo{pages}{034501} (\bibinfo{year}{2019}).
 
 
\bibitem{molliere2019}
\bibinfo{author}{{Molli{\`e}re}, P.} \emph{et~al.}
\newblock \bibinfo{title}{{petitRADTRANS. A Python radiative transfer package for exoplanet characterization and retrieval}}.
\newblock \emph{\bibinfo{journal}{\aap}}
\textbf{\bibinfo{volume}{627}}, \bibinfo{pages}{A67} (\bibinfo{year}{2019}).


\bibitem{tremblin2015}
\bibinfo{author}{{Tremblin}, P.} \emph{et~al.}
\newblock \bibinfo{title}{{Fingering Convection and Cloudless Models for Cool Brown Dwarf Atmospheres}}.
\newblock \emph{\bibinfo{journal}{\apjl}}
  \textbf{\bibinfo{volume}{804}}, \bibinfo{pages}{L17} (\bibinfo{year}{2015}).


\bibitem{baudino2015}
\bibinfo{author}{Baudino, J.-L., B{\'e}zard, B., Boccaletti, A., Bonnefoy, M., Lagrange, A.-M. \& Galicher, R.}
\newblock \bibinfo{title}{Interpreting the photometry and spectroscopy of directly imaged planets: a new atmospheric model applied to {\ensuremath{\beta}} Pictoris b and SPHERE observations}.
\newblock \emph{\bibinfo{journal}{\aap}}
  \textbf{\bibinfo{volume}{582}}, \bibinfo{pages}{A83} (\bibinfo{year}{2015}).

\bibitem{pandexo}
\bibinfo{author}{{Batalha}, N. E.} \emph{et~al.}
\newblock \bibinfo{title}{{PandExo: A Community Tool for Transiting Exoplanet Science with JWST \& HST}}.
\newblock \emph{\bibinfo{journal}{\pasp}}
\textbf{\bibinfo{volume}{129}}, \bibinfo{pages}{064501} (\bibinfo{year}{2017}).

\bibitem{schlawin2020}
\bibinfo{author}{{Schlawin}, E.} \emph{et~al.}
\newblock \bibinfo{title}{{JWST Noise Floor. I. Random Error Sources in JWST NIRCam Time Series}}.
\newblock \emph{\bibinfo{journal}{\aj}}
\textbf{\bibinfo{volume}{160}}, \bibinfo{pages}{231} (\bibinfo{year}{2020}).

\bibitem{sousa2020}
\bibinfo{author}{{Sousa-Silva}, C.} \emph{et~al.}
\newblock \bibinfo{title}{{Phosphine as a Biosignature Gas in Exoplanet Atmospheres}}.
\newblock \emph{\bibinfo{journal}{Astrobiology}}
 \textbf{\bibinfo{volume}{20}}, \bibinfo{pages}{235--268} (\bibinfo{year}{2020}).


\bibitem{lincowski2018}
\bibinfo{author}{{Lincowski}, A.} \emph{et~al.}
\newblock \bibinfo{title}{{Evolved Climates and Observational Discriminants for the TRAPPIST-1 Planetary System}}.
\newblock \emph{\bibinfo{journal}{\apj}}
  \textbf{\bibinfo{volume}{867}}, \bibinfo{pages}{76} (\bibinfo{year}{2018}).



\bibitem{li2015}
\bibinfo{author}{{Li}, G.} \emph{et~al.}
\newblock \bibinfo{title}{{Rovibrational Line Lists for Nine Isotopologues of the CO Molecule in the X $^{1}${\ensuremath{\Sigma}}$^{+}$ Ground Electronic State}}.
\newblock \emph{\bibinfo{journal}{\apjs}}
\textbf{\bibinfo{volume}{216}}, \bibinfo{pages}{15} (\bibinfo{year}{2015}).

\bibitem{wolniewicz1998}
\bibinfo{author}{Wolniewicz, L., Simbotin, I. \& Dalgarno, A.} 
\newblock \bibinfo{title}{{Quadrupole Transition Probabilities for the Excited Rovibrational States of H$_{2}$}}.
\newblock \emph{\bibinfo{journal}{\apjs}}
\textbf{\bibinfo{volume}{115}}, \bibinfo{pages}{293--313} (\bibinfo{year}{1998}).


\bibitem{polyansky2018}
\bibinfo{author}{{Polyansky}, O. L.} \emph{et~al.}
\newblock \bibinfo{title}{{ExoMol molecular line lists XXX: a complete high-accuracy line list for water}}.
\newblock \emph{\bibinfo{journal}{\mnras}}
\textbf{\bibinfo{volume}{480}}, \bibinfo{pages}{2597--2608} (\bibinfo{year}{2018}).


\bibitem{yurchenko2020}
\bibinfo{author}{Yurchenko, S. N., Mellor, T. M., Freedman, R. S. \& Tennyson, J.} 
\newblock \bibinfo{title}{{ExoMol line lists - XXXIX. Ro-vibrational molecular line list for CO$_{2}$}}.
\newblock \emph{\bibinfo{journal}{\mnras}}
\textbf{\bibinfo{volume}{496}}, \bibinfo{pages}{5282--5291} (\bibinfo{year}{2020}).


\bibitem{yurchenko2014}
\bibinfo{author}{Yurchenko, S. N. \& Tennyson, J. } 
\newblock \bibinfo{title}{{ExoMol line lists - IV. The rotation-vibration spectrum of methane up to 1500 K}}.
\newblock \emph{\bibinfo{journal}{\mnras}}
\textbf{\bibinfo{volume}{440}}, \bibinfo{pages}{1649--1661} (\bibinfo{year}{2014}).


\bibitem{yurchenko2017}
\bibinfo{author}{Yurchenko, S. N., Amundsen, D. S., Tennyson, J. \& Waldmann, I. P.} 
\newblock \bibinfo{title}{{A hybrid line list for CH$_{4}$ and hot methane continuum}}.
\newblock \emph{\bibinfo{journal}{\aap}}
\textbf{\bibinfo{volume}{605}}, \bibinfo{pages}{A95} (\bibinfo{year}{2017}).


\bibitem{roueff2019}
\bibinfo{author}{{Roueff}, E.} \emph{et~al.}
\newblock \bibinfo{title}{{The full infrared spectrum of molecular hydrogen}}.
\newblock \emph{\bibinfo{journal}{\aap}}
\textbf{\bibinfo{volume}{630}}, \bibinfo{pages}{A58} (\bibinfo{year}{2019}).


\end{thebibliography}

\newpage

\renewcommand{\figurename}{Supplementary\,figure}

\section{Tables and Figures for Methods}


\begin{figure*}[ht!]
    \centering
    \includegraphics[width=0.95\textwidth, trim = 0cm 1.5cm 0cm 1cm, clip=true]{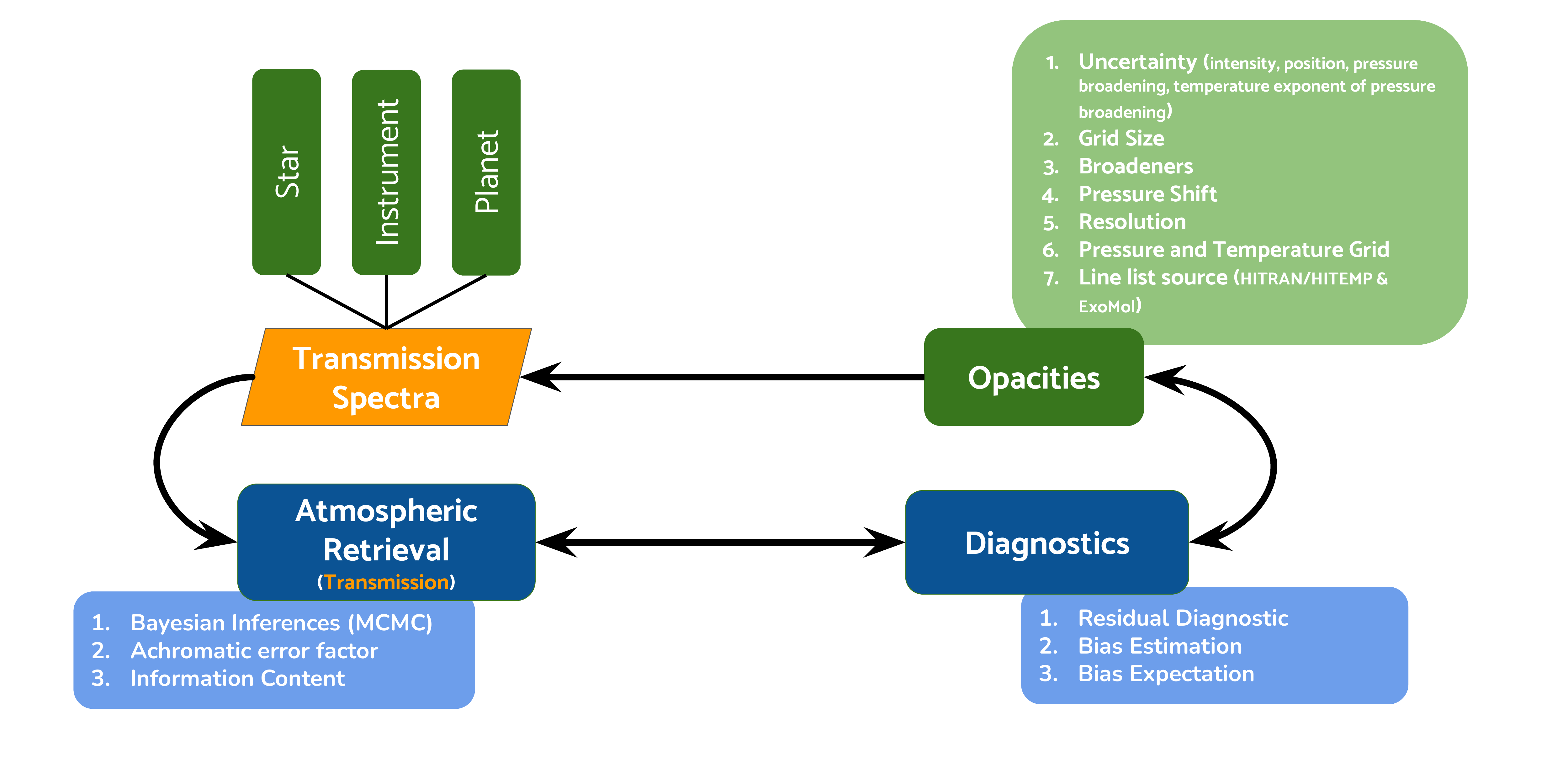}
    
    \caption{\label{fig:framework} \textbf{Framework for sensitivity analysis of retrieved atmospheric properties to opacity model.} The four building blocks of the transmission spectra are shown in green, analysis techniques are marked in blue, and the orange parallelogram highlights the remote sensing technique at the center of this perturbation/sensitivity analysis. The arrows are shown to indicate the direction of the information flow. We perform self- and cross-retrieval for two planetary cases (a super-Earth around M-dwarf and a Jupiter-sized planet around K-dwarf, see Supplementary Table\,\ref{tab:PlanetParams}) with nine distinct sets of cross-sections (see Supplementary Table\,\ref{tab:CS_parameter}) to access their impacts.}
    
\end{figure*}

\begin{table}
    \centering
    \caption{Parameters and line list sources used in the generation of our set of nine different cross-sections}
    \resizebox{\textwidth}{!}{
    
    \begin{tabular}{rccccccccc}
         \textbf {Profile:} & \multicolumn{9}{l}{ Voigt}\\
         \textbf{Resolution:} & \multicolumn{9}{l}{ 100,000 } \\
         \textbf{Wavelength Range:} & \multicolumn{9}{l}{0.6 $\mu$m to 30  $\mu$m}\\
         \textbf{Molecules:} & \multicolumn{8}{l}{H$_{\rm 2}$O, CO, CO$_{2}$, O$_3$, CH$_{\rm 4}$, N$_{\rm 2}$,  H$_{\rm 2}$}\\
        \hhline{==========}
        \multicolumn{9}{l}{\textbf{Line List Source}}\\
        \hline
        \textbf{HITRAN:} & \multicolumn{9}{l}{ H$_{\rm 2}$O [G17], CO [L15, G17], CO$_{2}$ [G17], O$_3$ [G17], CH$_{\rm 4}$ [G17], N$_{\rm 2}$ [G17],  H$_{\rm 2}$ [W98, G17]}\\
        \textbf{ExoMol:} & \multicolumn{9}{l}{ H$_{\rm 2}$O [P18], CO [L15], CO$_{2}$ [Y20], CH$_{\rm 4}$ [Y14, Y17],  H$_{\rm 2}$ [R19]}\\
        \textbf{HITEMP:} & \multicolumn{9}{l}{CH$_{\rm 4}$ [H20], H$_{\rm 2}$O [R10], CO$_{\rm 2}$ [R10], CO [L15]}\\
        \textbf{Rayleigh:} & \multicolumn{8}{l}{ H$_{\rm 2}$O [SU05], CO [SU05], CO$_{2}$ [SU05], CH$_{\rm 4}$ [SU05], N$_{\rm 2}$ [SU05],  H$_{\rm 2}$ [D65]}\\
        \textbf{CIA:}&\multicolumn{8}{l}{H$_{\rm 2}$-H$_{\rm 2}$ [A11], H$_{\rm 2}$-He [A12], N$_{\rm 2}$-N$_{\rm 2}$ [L96, K15, H17]} \\
        \hhline{==========}
       \multicolumn{9}{l}{\textbf{Grid}}\\
        \hline
        \textbf{Temperature (K):}&\multicolumn{9}{l}{[100, 110, 120, 130, 140, 160, 180, 200, 230, 260, 290, 330, 370, 410, 460, 510, 580, 650, 730, 810]}\\
        \textbf{Log$_{10}$ Pressure (atm):}&\multicolumn{9}{l}{[ -5.00, -4.20, -3.80,  -3.50, -3.25, -3.05, -2.95, -2.85, -2.75, -2.65, -2.55, -2.45, -2.30,  -2.15, -2.00,  -1.85, } \\
        &\multicolumn{9}{l}{-1.70, -1.55, -1.4,  -1.25, -1.1,  -0.95, -0.80,  -0.70, -0.60,  -0.50,  -0.40,  -0.30,  -0.20,  \-0.10,  0.00,    0.10, 0.20,  }\\
        &\multicolumn{9}{l}{ 0.30,   0.40,   0.50,   0.60,   0.70,   0.80,   0.90, 1.00,    1.10,   1.20,   1.30,   1.40,  1.50,   1.60,   1.70,   1.80,   1.90,   2.00]}\\
        \hhline{==========}
        \multicolumn{10}{l}{\textbf{Individual Parameters}}\\
        \hline
        & \textbf{CS-DFLT} &  \textbf{CS-1SUP} & \textbf{CS-1SDN} & \textbf{CS-SELF} & \textbf{CS-500W} & \textbf{CS-MAXB} & \textbf{CS-MINB} & \textbf{CS-EXML} & \textbf{CS-HTMP}\\
        \hline
        \textbf{Values} & Center & +1$\sigma ^\dagger$ & -1$\sigma ^\dagger$ & Center & Center & Center & Center & Center & Center \\
        \textbf{Broadening} ($\gamma$) & Air ($\gamma_A$) & $\gamma_A$ & $\gamma_A$ & Self ($\gamma_s$) & $\gamma_A$ & 2$\cdot$max($\gamma_{A}, \gamma_{S}$) & $\frac{1}{2}\cdot$ min($\gamma_{A}, \gamma_{S}$) & $\gamma$=0.07, n$^\ddagger$=0.5 & $\gamma_A$  \\
        \textbf{Linewing} (HWHM) & 50 & 50 & 50 & 50 & 500 & 50 & 50 & 50 & 50\\
        \textbf{Database} & HITRAN & HITRAN & HITRAN & HITRAN & HITRAN & HITRAN & HITRAN & ExoMol   & HITEMP \\
        \hline
        \end{tabular}
        }
        {\scriptsize \raggedright \textbf{References:} G17\cite{hitran2016}, L15\cite{li2015}, W98\cite{wolniewicz1998}, P18\cite{polyansky2018}, Y20\cite{yurchenko2020}, Y14\cite{yurchenko2014}, Y17\cite{yurchenko2017}, R19\cite{roueff2019}, SU05\cite{sneep2005}, D65\cite{dalgarno1965}, H20\cite{hargreaves2020}, R10\cite{HITEMP2010}, A11\cite{abel2011},  A12\cite{abel2012}, L96\cite{lafferty1996}, K15\cite{karman2015}, H17\cite{hartmann2017}. }\\
        {\scriptsize \raggedright $\dagger$Perturbed line position, line intensity, pressure broadening, and temperature exponent of the pressure broadening. \par}
        {\scriptsize \raggedright $\ddagger$Temperature exponent dependence of the pressure broadening. \par}
        \label{tab:CS_parameter}
\end{table}

\begin{figure*}[ht!]
    \centering
    \includegraphics[width=\textwidth, trim = 0cm 0cm 0cm 0cm, clip=true]{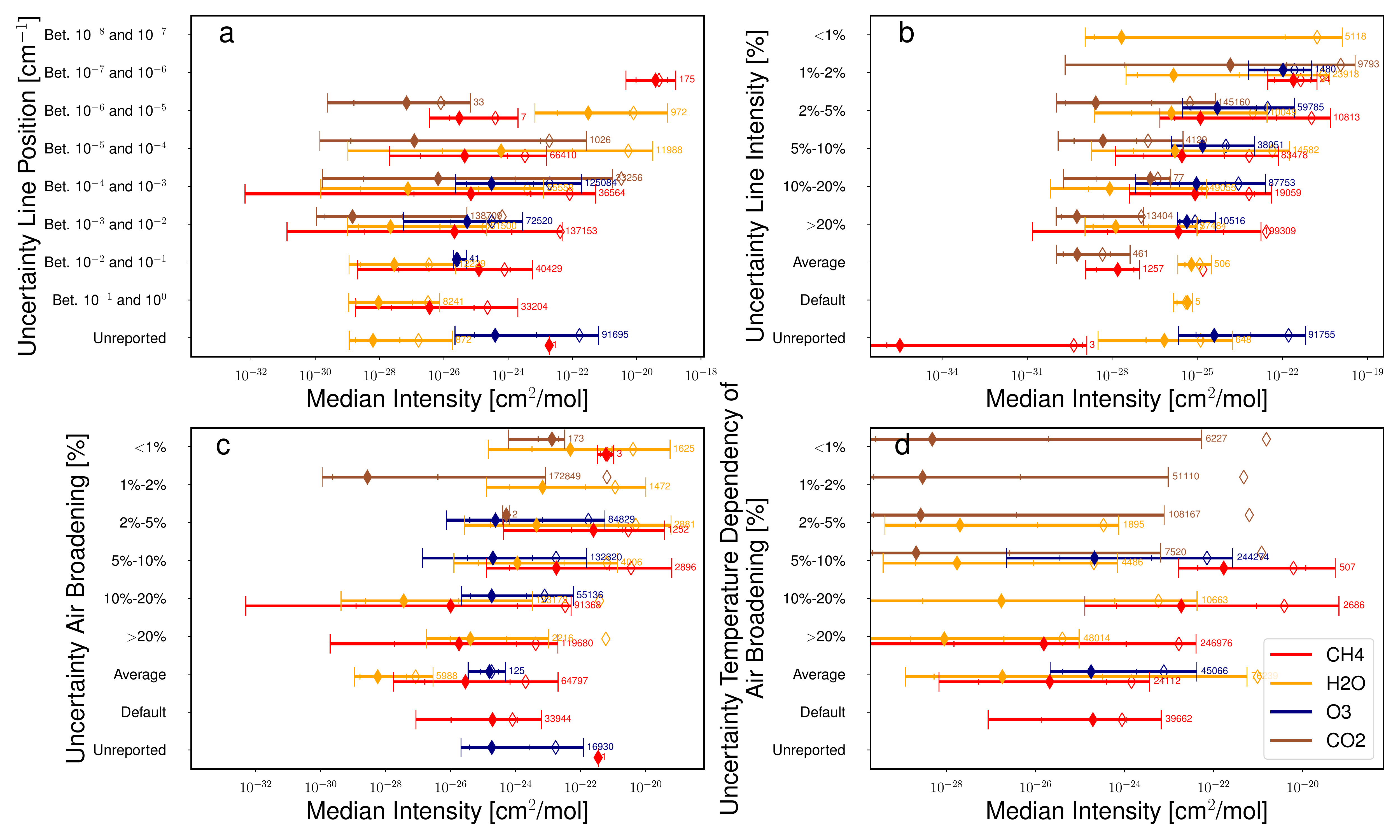}
    \caption{\label{fig:lineparamuncertainty} \textbf{Measurement uncertainties on absorption line parameters reported in \texttt{HITRAN}.} Uncertainties on the line parameters vs line intensities as reported in \texttt{HITRAN} for methane (red),  water (yellow), ozone (blue), and carbon dioxide (brown). For each uncertainty range reported (y axis), the mean and median line intensity values are shown as empty and full diamonds, with the 1 and 2$\sigma$ intensity ranges. The number of lines in each uncertainty range is reported on the right side of its 2$\sigma$ interval. These uncertainties are used to perturb the four different parameters in the generation of  \texttt{CS-1SUP} and \texttt{CS-1SDN}. The parameters of the strongest lines are reported with the smallest uncertainties. \textbf{a:} Uncertainty on the line position. \textbf{b:} Uncertainty on the  line intensity. \textbf{c:} Uncertainty on the air broadening coefficient. \textbf{d:} Uncertainty on the temperature dependency of the air broadening coefficient. The uncertainty codes (average, default, unreported) are introduced in Ref.\cite{hitran2004}.}
\end{figure*}

\begin{figure*}[ht]
\centering
   \includegraphics[width=.95\textwidth, trim=0cm 0cm 0.2cm 0.5cm, clip=true]{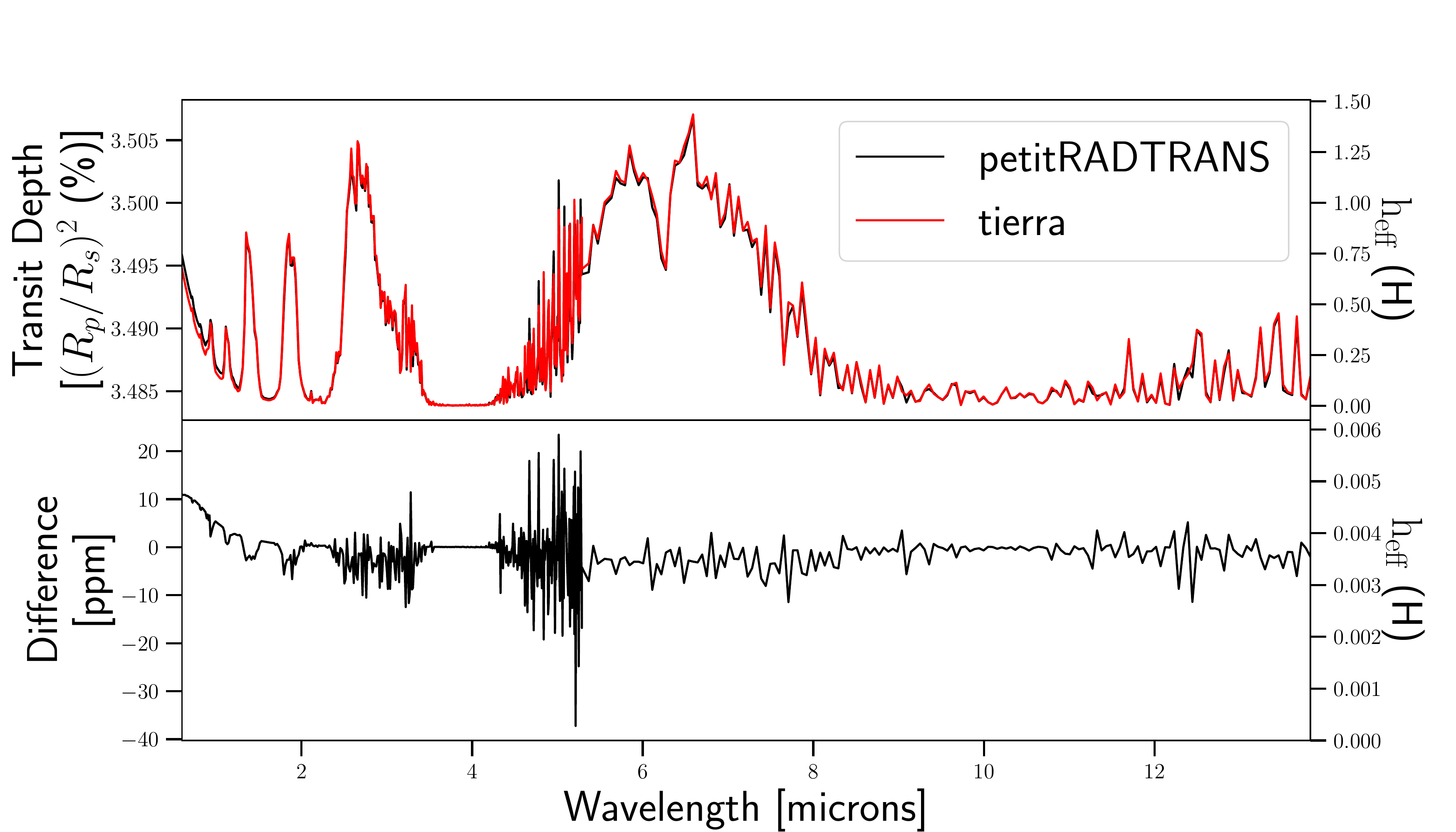}
   \caption{\label{fig:BenchmarkTest} \textbf{Benchmarking of \texttt{tierra} against \texttt{petitRadtrans}.} Benchmarking of \texttt{tierra} (R=100,000) against \texttt{petitRadtrans} for a Jupiter-sized planet around a K dwarf (0.55 R$_{\odot}$) for 365 K isothermal temperature with base pressure of 1 atm containing water at mixing ratio of $10^{-5}$ for a combined NIRSpec and MIRI observation. We use the  cross-sections from petitRadtrans, which present a lower resolution on their temperature and pressure grids, to focus this benchmarking on \texttt{tierra} rather than possible difference on underlying cross-sections. The median absolute deviation between two models is 1.95 p.p.m. and RMS is 4.4 ppm  which is marginal in comparison to the deviations seen in the model comparisons (Figure\,\ref{fig:ModelComparison}).}
\end{figure*}

\newpage


\begin{table}
\caption{\label{tab:PlanetParams}Parameters used for the generation of synthetic model }
\centering
\begin{tabular}{llll}
\multicolumn{4}{l}{\textbf{super-Earth}}\\
\hline
\hhline{====}
$\rm M_{Pl}$ & 1 M$_{\oplus}$ & T$_0$ & 300 K  \\
$\rm R_{Pl}$ & 1.1 R$_{\oplus}$& T$_\infty$ & 400 K \\
$\rm M_{*}$ & 0.1 M$_{\odot}$ & H$_{\rm T}$ & 10$^{-2}$ km$^{-1}$  \\
$\rm R_{*}$ & 0.1 M$_{\odot}$ & H-Mag & 10.0 \\
\hline
MR$_{\rm N_2}$ & 3.0$\times 10^{-1}$ & MR$_{\rm CH_4}$ & 1.0$\times 10^{-1}$\\
MR$_{\rm CO_2}$ & 1.0$\times 10^{-3}$ & MR$_{\rm O_3}$ &1.0$\times 10^{-4}$\\
MR$_{\rm CO}$ & 1.0$\times 10^{-1}$ & MR$_{\rm H_2}$ & 3.391$\times 10^{-1}$\\
MR$_{\rm H_2O}$ & 1.0$\times 10^{-1}$&  P$_{0}$ & 1.0 atm\\
\hline
\multicolumn{4}{l}{\textbf{Warm-Jupiter}}\\
\hline
$\rm M_{Pl}$ & 1.0 M$_{\rm Jup}$ & T$_0$ & 500 K  \\
$\rm R_{Pl}$ & 1.0 R$_{\rm Jup}$& T$_\infty$ & 600 K \\
$\rm M_{*}$ & 0.55 M$_{\odot}$ & H$_{\rm T}$ & 10$^{-2}$ km$^{-1}$  \\
$\rm R_{*}$ & 0.55 M$_{\odot}$ & H-Mag & 10.0 \\
\hline
MR$_{\rm N_2}$ & 1.0$\times 10^{-5}$& MR$_{\rm CH_4}$ & 2.0$\times 10^{-5}$ \\
MR$_{\rm CO_2}$ & 2.0$\times 10^{-5}$& MR$_{\rm O_3}$ & 1.0$\times 10^{-6}$ \\
MR$_{\rm CO}$ & 1.0$\times 10^{-5}$& MR$_{\rm H_2}$ & 8.499$\times 10^{-1}$ \\
MR$_{\rm H_2O}$ & 2.0$\times 10^{-5}$ &  P$_{0}$ & 1.0 atm\\
\hline
\end{tabular}

\end{table}

\begin{table}
\caption{Retrieved parameters$\dagger$ and their observed biases$\ddagger$ for the case of super-Earth}
\resizebox{\textwidth}{!}{
\centering
\begin{tabular}{r|c|c|c|c|c|c|c|c|c}
\hline
&\textbf{CS-DFLT}&\textbf{CS-1SUP}&\textbf{CS-1SDN}&\textbf{CS-SELF}&\textbf{CS-500W}&\textbf{CS-MAXB}&\textbf{CS-MINB}&\textbf{CS-EXML}&\textbf{CS-HTMP}\\
\hhline{==========}
T$_0$ [K]& $286.5^{+24.63}_{-45.66}$ & $206.29^{+50.94}_{-56.13}$ & $337.84^{+7.47}_{-13.2}$ & $186.53^{+28.22}_{-29.98}$ & $147.26^{+23.07}_{-20.22}$ & $130.66^{+25.6}_{-19.49}$ & $357.93^{+7.53}_{-9.36}$ & $110.21^{+10.44}_{-7.67}$& $107.67^{+9.93}_{-5.39}$ \\
& -0.55$\sigma$  & -1.84$\sigma$ $\vert$ 1.82 & \textbf{2.87}$\mathbf{\sigma}$ $\vert$ \textbf{0.35} & \textbf{-4.02}$\mathbf{\sigma}$ $\vert$ 0.99 & \textbf{-6.62}$\mathbf{\sigma}$ $\vert$ 0.73 & \textbf{-6.61}$\mathbf{\sigma}$ $\vert$ 0.77 & \textbf{6.19}$\mathbf{\sigma}$ $\vert$ \textbf{0.29} & \textbf{-18.18}$\mathbf{\sigma}$ $\vert$ \textbf{0.31}& \textbf{-19.37}$\mathbf{\sigma}$ $\vert$ \textbf{0.22} \\
\hline
Log$_{10}$H$_{\rm{T}}$ [m$^{-1}$] & $2.05^{+0.24}_{-0.21}$ & $-1.7^{+0.26}_{-0.18}$ & $-2.94^{+0.49}_{-0.3}$ & $-1.84^{+0.09}_{-0.09}$ & $-1.84^{+0.05}_{-0.06}$ & $-1.79^{+0.05}_{-0.07}$ & $-3.01^{+0.43}_{-0.46}$ & $0.11^{+0.63}_{-0.86}$& $0.29^{+0.47}_{-0.49}$ \\
& -0.21$\sigma$ & 1.67$\sigma$ $\vert$ 1.38 & -1.92$\sigma$ $\vert$ \textbf{2.47} & 1.78$\sigma$ $\vert$ 0.56 & \textbf{2.67}$\mathbf{\sigma}$ $\vert$ \textbf{0.34} & \textbf{3.0}$\mathbf{\sigma}$ $\vert$ \textbf{0.38} & -2.35$\sigma$ $\vert$ \textbf{2.78} & 2.45$\sigma$ $\vert$ \textbf{4.66}& \textbf{4.67}$\mathbf{\sigma}$ $\vert$ \textbf{2.13} \\
\hline
T$_\infty$ [K]& $412.21^{+16.83}_{-10.86}$ & $399.21^{+9.94}_{-6.14}$ & $553.88^{+166.78}_{-112.72}$ & $405.76^{+8.22}_{-7.18}$ & $421.64^{+4.48}_{-3.95}$ & $408.89^{+7.25}_{-5.63}$ & $531.03^{+210.67}_{-90.37}$ & $370.75^{+1.51}_{-1.89}$& $371.36^{+1.88}_{-1.87}$ \\
&\blue{1.12$\sigma$} & -0.08$\sigma$ $\vert$ 0.83 & 1.37$\sigma$ $\vert$ \textbf{14.49} & 0.8$\sigma$ $\vert$ 0.8 & \textbf{5.48}$\mathbf{\sigma}$ $\vert$ \textbf{0.44} & 1.58$\sigma$ $\vert$ 0.67 & 1.45$\sigma$ $\vert$ \textbf{15.61} & \textbf{-19.37}$\mathbf{\sigma}$ $\vert$ \textbf{0.18}& \blue{\textbf{-15.23}$\mathbf{\sigma}$ $\vert$ \textbf{0.14}} \\
\hline
Log$_{10}$N$_{\rm N_2}$ [m$^{-3}$]& \blue{$23.24^{+0.13}_{-0.13}$} & $23.26^{+0.28}_{-0.35}$ & $23.08^{+0.08}_{-0.11}$ & $23.19^{+0.29}_{-0.33}$ & $11.74^{+7.1}_{-7.77}$ & $23.12^{+0.38}_{-0.72}$ & $23.17^{+0.08}_{-0.13}$ & $23.24^{+0.04}_{-0.04}$& \blue{$23.27^{+0.06}_{-0.07}$ } \\
&\blue{0.18$\sigma$} & 0.13$\sigma$ $\vert$ 1.66 & -1.7$\sigma$ $\vert$ 0.5 & -0.09$\sigma$ $\vert$ 1.63 & -1.62$\sigma$ $\vert$ \textbf{39.13} & -0.25$\sigma$ $\vert$ \textbf{2.89} & -0.58$\sigma$ $\vert$ 0.55 & 0.59$\sigma$ $\vert$ \textbf{0.21}& \blue{0.77$\sigma$ $\vert$ 0.5} \\
\hline
Log$_{10}$N$_{\rm CO}$ [m$^{-3}$]&\blue{$22.63^{+0.24}_{-0.31}$} & $23.05^{+0.23}_{-0.27}$ & $22.05^{+0.34}_{-0.44}$ & $23.08^{+0.25}_{-0.39}$ & $21.9^{+0.26}_{-0.27}$ & $23.12^{+0.33}_{-0.4}$ & $22.58^{+0.33}_{-0.37}$ & $22.03^{+0.26}_{-0.32}$& \blue{$22.96^{+0.04}_{-0.04}$} \\
&\blue{-0.45$\sigma$} & 1.15$\sigma$ $\vert$ 1.11 & -2.03$\sigma$ $\vert$ 1.73 & 0.87$\sigma$ $\vert$ 1.42 & \textbf{-3.23}$\mathbf{\sigma}$ $\vert$ 1.18 & 0.95$\sigma$ $\vert$ 1.62 & -0.48$\sigma$ $\vert$ 1.56 & \textbf{-2.73}$\mathbf{\sigma}$ $\vert$ 1.29& \blue{\textbf{5.52}$\mathbf{\sigma}$ $\vert$ \textbf{0.15}}\\
\hline
Log$_{10}$N$_{\rm H_2O}$ [m$^{-3}$]& \blue{$22.9^{+0.08}_{-0.09}$} & $22.93^{+0.09}_{-0.06}$ & $22.83^{+0.06}_{-0.06}$ & $23.07^{+0.11}_{-0.1}$ & $23.71^{+0.12}_{-0.12}$ & $23.37^{+0.1}_{-0.1}$ & $22.69^{+0.06}_{-0.06}$ & $22.24^{+0.06}_{-0.04}$& \blue{$22.12^{+0.03}_{-0.03}$} \\
&\blue{1.79$\sigma$} & \textbf{3.18}$\mathbf{\sigma}$ $\vert$ 0.79 & 1.52$\sigma$ $\vert$ 0.63 & \textbf{3.31}$\mathbf{\sigma}$ $\vert$ 1.11 & \textbf{8.09}$\mathbf{\sigma}$ $\vert$ 1.26 & \textbf{6.31}$\mathbf{\sigma}$ $\vert$ 1.05 & -0.82$\sigma$ $\vert$ 0.63 & \textbf{-8.32}$\mathbf{\sigma}$ $\vert$ 0.53& \blue{\textbf{-20.64}$\mathbf{\sigma}$ $\vert$ \textbf{0.35}}\\
\hline
Log$_{10}$N$_{\rm CO_2}$ [m$^{-3}$]&\blue{$20.88^{+0.11}_{-0.1}$} & $20.93^{+0.11}_{-0.09}$ & $20.82^{+0.09}_{-0.08}$ & $21.43^{+0.11}_{-0.1}$ & $21.79^{+0.13}_{-0.13}$ & $21.75^{+0.09}_{-0.11}$ & $20.68^{+0.1}_{-0.1}$ & $19.84^{+0.07}_{-0.07}$& \blue{$19.59^{+0.07}_{-0.07}$}\\
&\blue{1.41$\sigma$} & 2.12 $\vert$ 0.95 & 1.01$\sigma$ $\vert$ 0.81 & \textbf{6.91}$\mathbf{\sigma}$ $\vert$ 1.0 & \textbf{8.08}$\mathbf{\sigma}$ $\vert$ 1.24 & \textbf{9.19}$\mathbf{\sigma}$ $\vert$ 0.95 & -0.59$\sigma$ $\vert$ 0.95 & \textbf{-12.84}$\mathbf{\sigma}$ $\vert$ 0.67& \blue{\textbf{-16.42}$\mathbf{\sigma}$ $\vert$ 0.67} \\
\hline
Log$_{10}$N$_{\rm CH_4}$ [m$^{-3}$]&\blue{$22.88^{+0.07}_{-0.08}$} & $22.82^{+0.09}_{-0.06}$ & $22.94^{+0.06}_{-0.05}$ & $23.31^{+0.1}_{-0.08}$ & $23.75^{+0.11}_{-0.1}$ & $23.64^{+0.09}_{-0.08}$ & $22.65^{+0.06}_{-0.05}$ & $21.48^{+0.07}_{-0.02}$& \blue{$21.42^{+0.02}_{-0.02}$} \\
&\blue{1.76$\sigma$} & 1.35$\sigma$ $\vert$ 0.94 & \textbf{4.02}$\mathbf{\sigma}$ $\vert$ 0.69 & \textbf{7.14}$\mathbf{\sigma}$ $\vert$ 1.12 & \textbf{10.11}$\mathbf{\sigma}$ $\vert$ 1.31 & \textbf{11.26}$\mathbf{\sigma}$ $\vert$ 1.06 & -1.48$\sigma$ $\vert$ 0.69 & \textbf{-17.99}$\mathbf{\sigma}$ $\vert$ 0.56& \blue{\textbf{-65.95}$\mathbf{\sigma}$ $\vert$ \textbf{0.27}} \\
\hline
Log$_{10}$N$_{\rm O_3}$ [m$^{-3}$]&\blue{$19.92^{+0.09}_{-0.09}$} & $19.84^{+0.11}_{-0.07}$ & $19.94^{+0.08}_{-0.07}$ & $20.35^{+0.1}_{-0.1}$ & $20.82^{+0.12}_{-0.11}$ & $20.66^{+0.11}_{-0.09}$ & $19.66^{+0.09}_{-0.07}$ & $19.05^{+0.08}_{-0.05}$& \blue{$18.86^{+0.05}_{-0.05}$} \\
&\blue{2.01$\sigma$} & 1.44$\sigma$ $\vert$ 0.95 & \textbf{2.87}$\mathbf{\sigma}$ $\vert$ 0.79 & \textbf{6.11}$\mathbf{\sigma}$ $\vert$ 1.05 & \textbf{9.83}$\mathbf{\sigma}$ $\vert$ 1.21 & \textbf{10.23}$\mathbf{\sigma}$ $\vert$ 1.05 & -0.88$\sigma$ $\vert$ 0.84 & \textbf{-8.61}$\mathbf{\sigma}$ $\vert$ 0.68& \blue{ \textbf{-17.58}$\mathbf{\sigma}$ $\vert$ 0.56}\\
\hline
Log$_{10}$N$_{\rm H_2}$ [m$^{-3}$]&\blue{$23.25^{+0.12}_{-0.12}$} & $23.4^{+0.13}_{-0.13}$ & $23.03^{+0.08}_{-0.09}$ & $23.36^{+0.18}_{-0.18}$ & $11.75^{+6.75}_{-7.96}$ & $23.39^{+0.21}_{-0.37}$ & $23.17^{+0.05}_{-0.04}$ & $23.38^{+0.06}_{-0.02}$& \blue{$23.56^{+0.04}_{-0.04}$} \\
&\blue{-1.17$\sigma$} & 0.07$\sigma$ $\vert$ 1.3 & \textbf{-4.51}$\mathbf{\sigma}$ $\vert$ 0.85 & -0.17$\sigma$ $\vert$ 1.8 & -1.72$\sigma$ $\vert$ \textbf{73.55} & -0.0$\sigma$ $\vert$ \textbf{2.9} & \textbf{-4.41}$\mathbf{\sigma}$ $\vert$ \textbf{0.45} & -0.17$\sigma$ $\vert$ \textbf{0.4}& \blue{\textbf{4.24}$\mathbf{\sigma}$ $\vert$ \textbf{0.33}} \\
\hline
Red. $ \chi^2_{\nu}$ & 0.942 & 1.002 & 1.079 & 0.974 & 1.179 & 0.985 & 0.940 & 6.504& \blue{4.89} \\
SSD & - & \blue{276.76} & \blue{512.77} & \blue{983.69}  & \blue{1243.21} & \blue{2501.09} & \blue{247.28} & \blue{12624.09} & \blue{12035.61}\\
\blue{PSD} & - & \blue{4.74} & \blue{8.91} & \blue{39.47}  & \blue{33.68} & \blue{50.2} & \blue{5.79} & \blue{245.63} & \blue{383.04} \\
\hline
\end{tabular}
}
{\scriptsize $\dagger$ Retrieved parameters are introduced on the first line of a cell with its upper and lower 1$\sigma$ confidence interval. \par}
{\scriptsize $\ddagger$ Observed biases are introduced on the second line of a cell as the difference between the retrieved parameter and the nominal value divided by the retrieved uncertainty interval. The second number reported on this line is the ratio between retrieved uncertainty interval and nominal uncertainty interval to highlight an artificial amplification or reduction of the interval. Biases over 2.5$\sigma$ and interval amplification/reduction over 2/below 0.5 are shown in bold. \par}
\label{tab:RetrievedParams_SE}
\end{table}

\begin{figure*}[ht]
\centering
    {\vspace{-3cm}\includegraphics[width=0.95\textwidth, trim=1cm 1.5cm 1cm 0.5cm, clip=true]{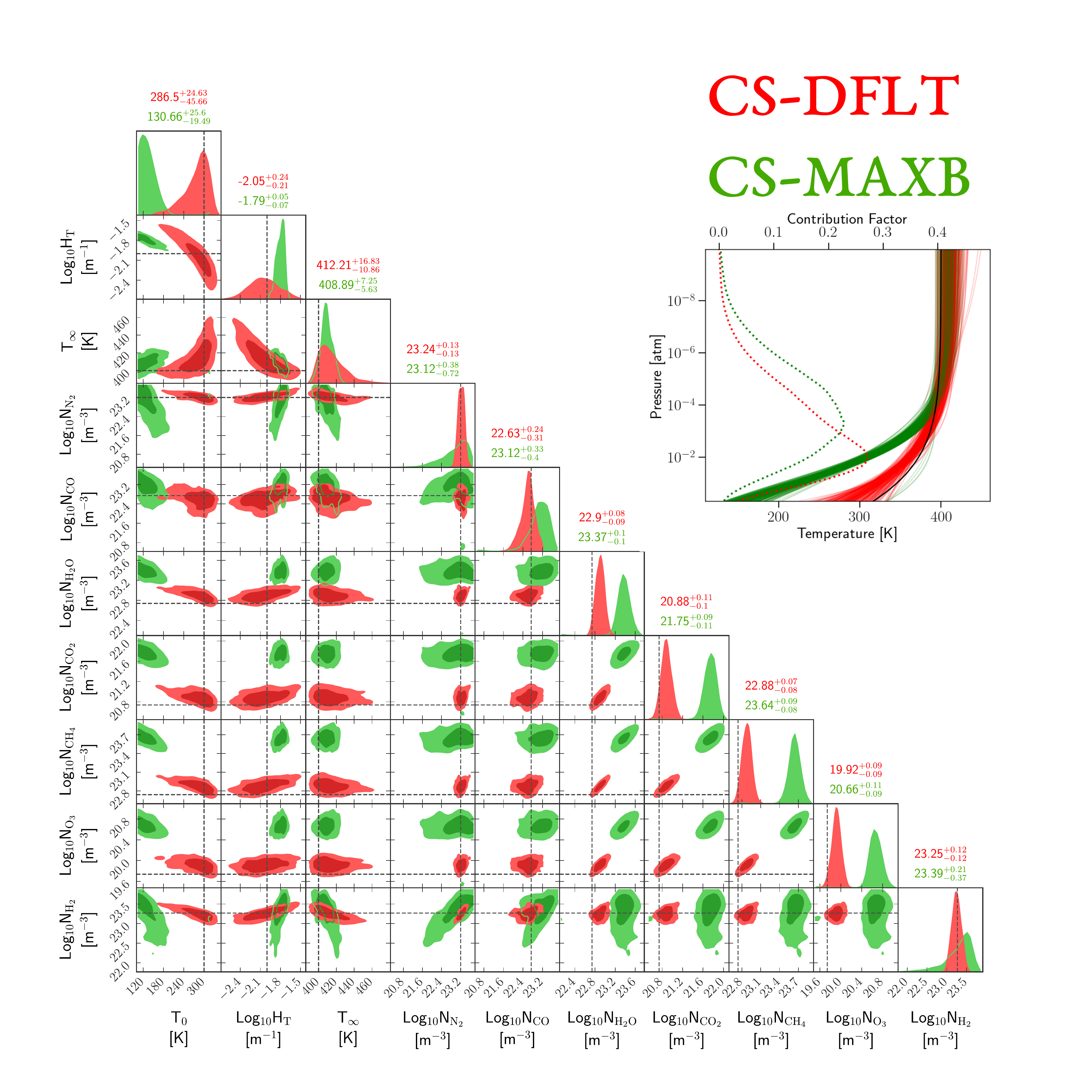}}
   \caption{\textbf{Corner plot of the posterior probability distribution of the atmospheric parameters for a super-Earth.} Corner plot showing the PPDs of the retrieved parameters for the case of the super-Earth for self-retrieval (\texttt{CS-DFLT}, red) and cross-retrieval (\texttt{CS-MAXB}, green). The only difference between the two cross-sections relate to broadening; \texttt{CS-DFLT} assumes air-broadening (geocentric) while  \texttt{CS-MAXB} assumes twice the maximum between air- and self-broadening. Strong biases are seen in the retrieved value of water (-6.97$\sigma$), carbon dioxide (-6.23$\sigma$), methane (-6.23$\sigma$), and ozone (-6.23$\sigma$). (Top Right) 500 random PT profiles constructed from the posteriors are shown for comparison against the true profile shown in black. The dotted line shows the contribution factor can change significantly due to the changes in the broadening values.}
   \label{fig:SE_CornerPlot}
\end{figure*}

\begin{table}
\caption{\label{tab:RetrievedParams_HJ}Retrieved parameters$\dagger$ and their observed biases$\ddagger$ for the case of warm-Jupiter}
\resizebox{\textwidth}{!}{
\centering
\begin{tabular}{r|c|c|c|c|c|c|c|c|c}
\hline
&\textbf{CS-DFLT}&\textbf{CS-1SUP}&\textbf{CS-1SDN}&\textbf{CS-SELF}&\textbf{CS-500W}&\textbf{CS-MAXB}&\textbf{CS-MINB}&\textbf{CS-EXML}& \blue{\textbf{CS-HTMP}}\\
\hhline{==========}
T$_0$ [K]&$418.94^{+66.68}_{-58.25}$ & $437.22^{+73.41}_{-61.21}$ & $418.88^{+69.2}_{-59.97}$ & $219.09^{+37.27}_{-32.61}$ & $200.77^{+37.91}_{-34.07}$ & $116.32^{+14.94}_{-10.63}$ & $674.88^{+100.0}_{-76.54}$ & $568.35^{+14.47}_{-50.67}$& \blue{$599.42^{+120.0}_{-88.46}$} \\
&-1.22$\sigma$ & -0.86$\sigma$ $\vert$ 1.08 & -1.17$\sigma$ $\vert$ 1.03 & \textbf{-7.54}$\mathbf{\sigma}$ $\vert$ 0.56 & \textbf{-7.89}$\mathbf{\sigma}$ $\vert$ 0.58 & \textbf{-25.68}$\mathbf{\sigma}$ $\vert$ \textbf{0.2} & 2.28$\sigma$ $\vert$ 1.41 & 1.35$\sigma$ $\vert$ 0.52& \blue{1.12$\sigma$ $\vert$ 1.67} \\
\hline
\blue{Log$_{10}$H$_{\rm T}$ [m$^{-1}$]}&$-1.75^{+0.38}_{-0.22}$ & $-1.66^{+1.15}_{-0.29}$ & $-1.7^{+1.11}_{-0.28}$ & $-1.57^{+0.26}_{-0.15}$ & $-1.13^{+0.92}_{-0.35}$ & $-1.5^{+0.2}_{-0.13}$ & $-0.9^{+1.3}_{-5.81}$ & $-3.21^{+1.89}_{-4.62}$& \blue{$-0.96^{+1.26}_{-0.97}$}  \\
&1.14$\sigma$ & 1.17$\sigma$ $\vert$ \textbf{2.4} & 1.07$\sigma$ $\vert$ \textbf{2.32} & \textbf{2.87}$\mathbf{\sigma}$ $\vert$ 0.68 & 2.49$\sigma$ $\vert$ \textbf{2.12} & \textbf{3.85}$\mathbf{\sigma}$ $\vert$ 0.55 & 0.19$\sigma$ $\vert$ \textbf{11.85} & -0.64$\sigma$ $\vert$ \textbf{10.85}& \blue{1.07$\sigma$ $\vert$ \textbf{3.72}}  \\
\hline
T$_\infty$ [K]&$602.77^{+6.41}_{-5.44}$ & $604.7^{+6.95}_{-5.72}$ & $599.2^{+8.11}_{-5.7}$ & $592.6^{+4.56}_{-4.52}$ & $599.15^{+4.72}_{-4.74}$ & $591.11^{+4.44}_{-4.21}$ & $600.22^{+21.54}_{-8.61}$ & $600.47^{+112.48}_{-22.18}$& \blue{$579.1^{+2.86}_{-3.46}$} \\
&0.51$\sigma$ & 0.82$\sigma$ $\vert$ 1.07 & -0.1$\sigma$ $\vert$ 1.17 & -1.62$\sigma$ $\vert$ 0.77 & -0.18$\sigma$ $\vert$ 0.8 & -2.0$\sigma$ $\vert$ 0.73 & 0.03$\sigma$ $\vert$ \textbf{2.54} & 0.02$\sigma$ $\vert$ \textbf{11.36}& \blue{\textbf{-7.31}$\mathbf{\sigma}$ $\vert$ 0.53}  \\
\hline
Log$_{10}$N$_{\rm N_2}$ [m$^{-3}$]&$9.98^{+6.85}_{-6.67}$ & $10.22^{+6.85}_{-7.04}$ & $11.18^{+6.96}_{-7.56}$ & $10.5^{+6.59}_{-6.98}$ & $9.96^{+7.75}_{-6.54}$ & $10.92^{+6.78}_{-7.36}$ & $10.92^{+6.73}_{-7.44}$ & $10.19^{+5.97}_{-6.52}$& \blue{$11.38^{+5.42}_{-6.69}$} \\
&-1.25$\sigma$ & -1.21$\sigma$ $\vert$ 1.03 & -1.05$\sigma$ $\vert$ 1.07 & -1.22$\sigma$ $\vert$ 1.0 & -1.1$\sigma$ $\vert$ 1.06 & -1.12$\sigma$ $\vert$ 1.05 & -1.13$\sigma$ $\vert$ 1.05 & -1.39$\sigma$ $\vert$ 0.92& \blue{-1.66$\sigma$ $\vert$ 1.09}  \\
\hline
Log$_{10}$N$_{\rm CO}$ [m$^{-3}$]&$19.32^{+0.26}_{-0.29}$ & $19.38^{+0.25}_{-0.28}$ & $19.17^{+0.25}_{-0.33}$ & $19.64^{+0.28}_{-0.3}$ & $18.42^{+0.19}_{-0.2}$ & $19.73^{+0.27}_{-0.28}$ & $19.19^{+0.29}_{-0.3}$ & $19.13^{+0.23}_{-0.3}$&\blue{$19.07^{+0.27}_{-0.33}$}  \\
&-0.76$\sigma$ & -0.55$\sigma$ $\vert$ 0.96 & -1.39$\sigma$ $\vert$ 1.05 & 0.41$\sigma$ $\vert$ 1.05 & \textbf{-5.77}$\mathbf{\sigma}$ $\vert$ 0.71 & 0.76$\sigma$ $\vert$ 1.0 & -1.13$\sigma$ $\vert$ 1.07 & -1.68$\sigma$ $\vert$ 0.96& \blue{-1.66$\sigma$ $\vert$ 1.09} \\
\hline
Log$_{10}$N$_{\rm H_2O}$ [m$^{-3}$]&$18.84^{+0.07}_{-0.06}$ & $18.81^{+0.08}_{-0.06}$ & $18.83^{+0.08}_{-0.07}$ & $18.72^{+0.08}_{-0.08}$ & $18.9^{+0.09}_{-0.07}$ & $18.78^{+0.09}_{-0.1}$ & $18.78^{+0.04}_{-0.04}$ & $18.85^{+0.06}_{-0.04}$& \blue{$18.84^{+0.07}_{-0.06}$}  \\
&0.36$\sigma$ & -0.1$\sigma$ $\vert$ 1.08 & 0.17$\sigma$ $\vert$ 1.15 & -1.23$\sigma$ $\vert$ 1.23 & 1.17$\sigma$ $\vert$ 1.23 & -0.43$\sigma$ $\vert$ 1.46 & -0.96$\sigma$ $\vert$ 0.62 & 0.79$\sigma$ $\vert$ 0.77& \blue{0.36$\sigma$ $\vert$ 1.0}  \\
\hline
Log$_{10}$N$_{\rm CO_2}$ [m$^{-3}$]&$18.54^{+0.07}_{-0.07}$ & $18.48^{+0.07}_{-0.07}$ & $18.58^{+0.08}_{-0.07}$ & $18.82^{+0.08}_{-0.09}$ & $18.62^{+0.09}_{-0.08}$ & $18.89^{+0.09}_{-0.1}$ & $18.49^{+0.06}_{-0.05}$ & $18.57^{+0.06}_{-0.05}$& \blue{$18.62^{+0.06}_{-0.06}$} \\
&0.33$\sigma$ & -0.53$\sigma$ $\vert$ 1.0 & 0.9$\sigma$ $\vert$ 1.07 & \textbf{3.37}$\mathbf{\sigma}$ $\vert$ 1.21 & 1.29$\sigma$ $\vert$ 1.21 & \textbf{3.73}$\mathbf{\sigma}$ $\vert$ 1.36 & -0.45$\sigma$ $\vert$ 0.79 & 1.06$\sigma$ $\vert$ 0.79& \blue{1.72$\sigma$ $\vert$ 0.86}  \\
\hline
Log$_{10}$N$_{\rm CH_4}$ (m$^{-3}$)&$18.82^{+0.06}_{-0.05}$ & $18.69^{+0.06}_{-0.05}$ & $18.94^{+0.07}_{-0.05}$ & $18.99^{+0.06}_{-0.08}$ & $18.86^{+0.08}_{-0.06}$ & $19.06^{+0.09}_{-0.1}$ & $18.74^{+0.03}_{-0.03}$ & $18.37^{+0.04}_{-0.03}$& \blue{$18.4^{+0.04}_{-0.03}$} \\
&0.04$\sigma$ & -2.13$\sigma$ $\vert$ 1.0 & 2.44$\sigma$ $\vert$ 1.09 & 2.15$\sigma$ $\vert$ 1.27 & 0.7$\sigma$ $\vert$ 1.27 & 2.42$\mathbf{\sigma}$ $\vert$ 1.73 & \textbf{-2.6}$\mathbf{\sigma}$ $\vert$ 0.55 & \textbf{-11.2}$\mathbf{\sigma}$ $\vert$ 0.64& \blue{\textbf{-10.45}$\mathbf{\sigma}$ $\vert$ 0.64}  \\
\hline
Log$_{10}$N$_{\rm O_3}$ (m$^{-3}$)&$17.55^{+0.07}_{-0.07}$ & $17.45^{+0.07}_{-0.06}$ & $17.66^{+0.08}_{-0.07}$ & $17.72^{+0.07}_{-0.08}$ & $17.63^{+0.08}_{-0.07}$ & $17.79^{+0.09}_{-0.1}$ & $17.49^{+0.05}_{-0.05}$ & $17.56^{+0.06}_{-0.06}$& \blue{$17.56^{+0.07}_{-0.06}$}  \\
&0.47$\sigma$ & -0.96$\sigma$ $\vert$ 0.93 & 2.04$\sigma$ $\vert$ 1.07 & \textbf{2.54}$\mathbf{\sigma}$ $\vert$ 1.07 & 1.61$\sigma$ $\vert$ 1.07 & \textbf{2.73}$\mathbf{\sigma}$ $\vert$ 1.36 & -0.54$\sigma$ $\vert$ 0.71 & 0.72$\sigma$ $\vert$ 0.86& \blue{0.72$\sigma$ $\vert$ 0.93} \\
\hline
Log$_{10}$N$_{\rm H_2}$ (m$^{-3}$)&$23.45^{+0.04}_{-0.05}$ & $23.42^{+0.05}_{-0.04}$ & $23.45^{+0.06}_{-0.05}$ & $23.56^{+0.05}_{-0.07}$ & $23.45^{+0.08}_{-0.05}$ & $23.64^{+0.08}_{-0.09}$ & $23.36^{+0.01}_{-0.01}$ & $23.42^{+0.03}_{-0.02}$& \blue{$23.42^{+0.03}_{-0.02}$}  \\
&-0.53$\sigma$ & -1.02$\sigma$ $\vert$ 1.0 & -0.35$\sigma$ $\vert$ 1.22 & 1.27$\sigma$ $\vert$ 1.33 & -0.26$\sigma$ $\vert$ 1.44 & 1.88$\sigma$ $\vert$ 1.89 & \textbf{-11.1}$\mathbf{\sigma}$ $\vert$ \textbf{0.22} & -1.7$\sigma$ $\vert$ 0.56& \blue{-1.7$\sigma$ $\vert$ 0.56}  \\
\hline
Red. $\chi^2_{\nu}$ & 1.038 & 1.065 & 1.031 & 1.068 & 1.076 & 1.069 & 1.046 & 1.996& \blue{1.788} \\
SSD & \blue{$-$} & \blue{89.79} & \blue{139.84} & \blue{613.77} & \blue{93.83} & \blue{1709.87} & \blue{119.44} & \blue{3655.73} & \blue{3536.44} \\
PSD & - & 4.90 & 7.88 & 19.37 & 6.44  & 27.67 & 6.56 & 108.42 & \blue{116.45} \\
\hline
\end{tabular}
}
{\scriptsize $\dagger$ Retrieved parameters are introduced on the first line of a cell with its upper and lower 1 sigma confidence interval. \par}
{\scriptsize $\ddagger$ Observed biases are introduced on the second line of a cell as the difference between the retrieved parameter and the nominal value divided by the retrieved uncertainty interval. The second number reported on this line is the ratio between retrieved uncertainty interval and nominal uncertainty interval to highlight an artificial amplification or reduction of the interval. Biases over 2.5$\sigma$ and interval amplification/reduction over 2/below 0.5 are shown in bold. \par}
\end{table}

\begin{figure*}[ht]
\centering
   \includegraphics[width=\textwidth, trim=3.5cm 1cm 1cm 0.5cm]{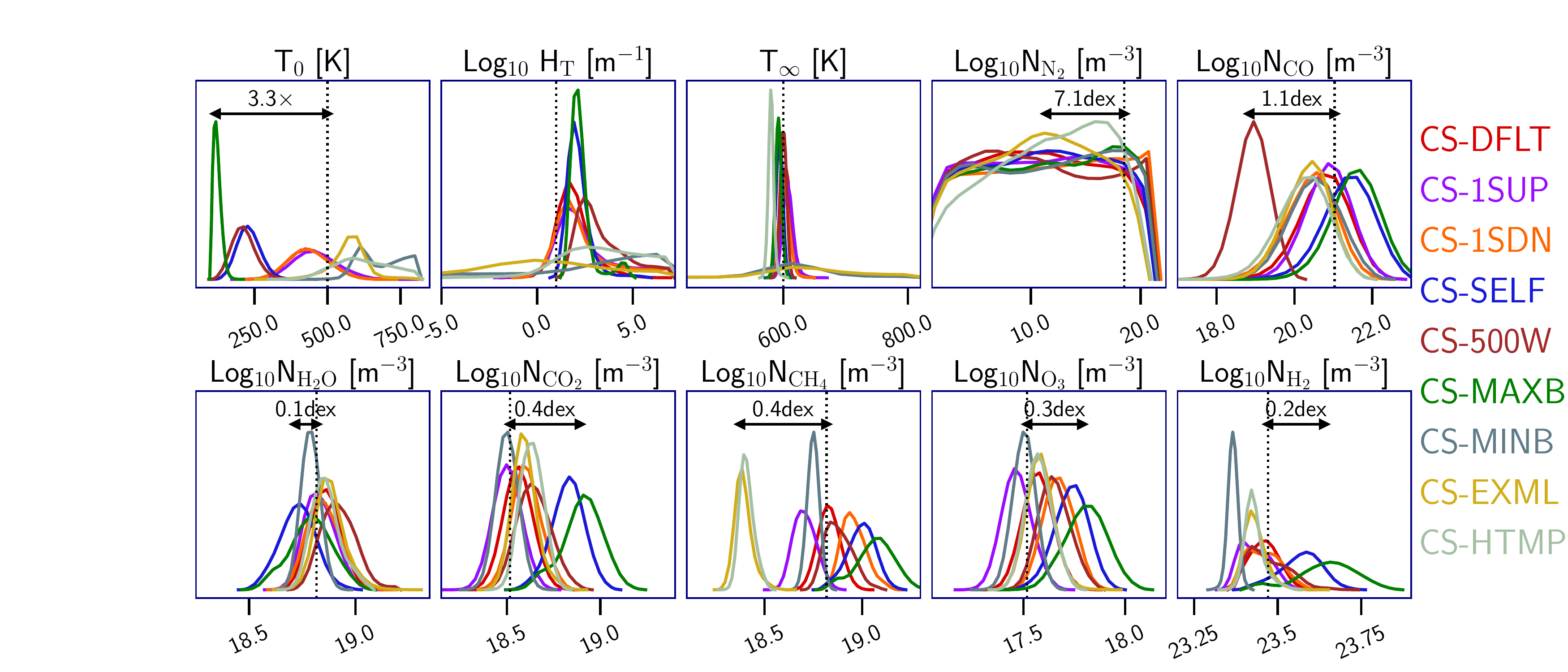} 
   \caption{\label{fig:HJParameters} \textbf{Propagation of the ensemble of opacity-model perturbations to the level of retrieved atmospheric properties for the warm-Jupiter scenario.} Posterior probability distributions (PPDs) of the retrieved atmospheric parameters for the warm-Jupiter scenario highlighting the biases induced by perturbations to the opacity model (see Methods). Each cross-section is identified by its color and label on the right. The dotted black vertical lines represent the true values used in generating the synthetic spectrum. Deviations with a statistical significance of up to 20$\sigma$ and physical significance of over 1~dex are reported.}
\end{figure*}

\begin{figure*}[ht]
\centering
   {\vspace{-3cm}\includegraphics[width=0.95\textwidth, trim=0cm 1cm 0cm 0cm, clip=true]{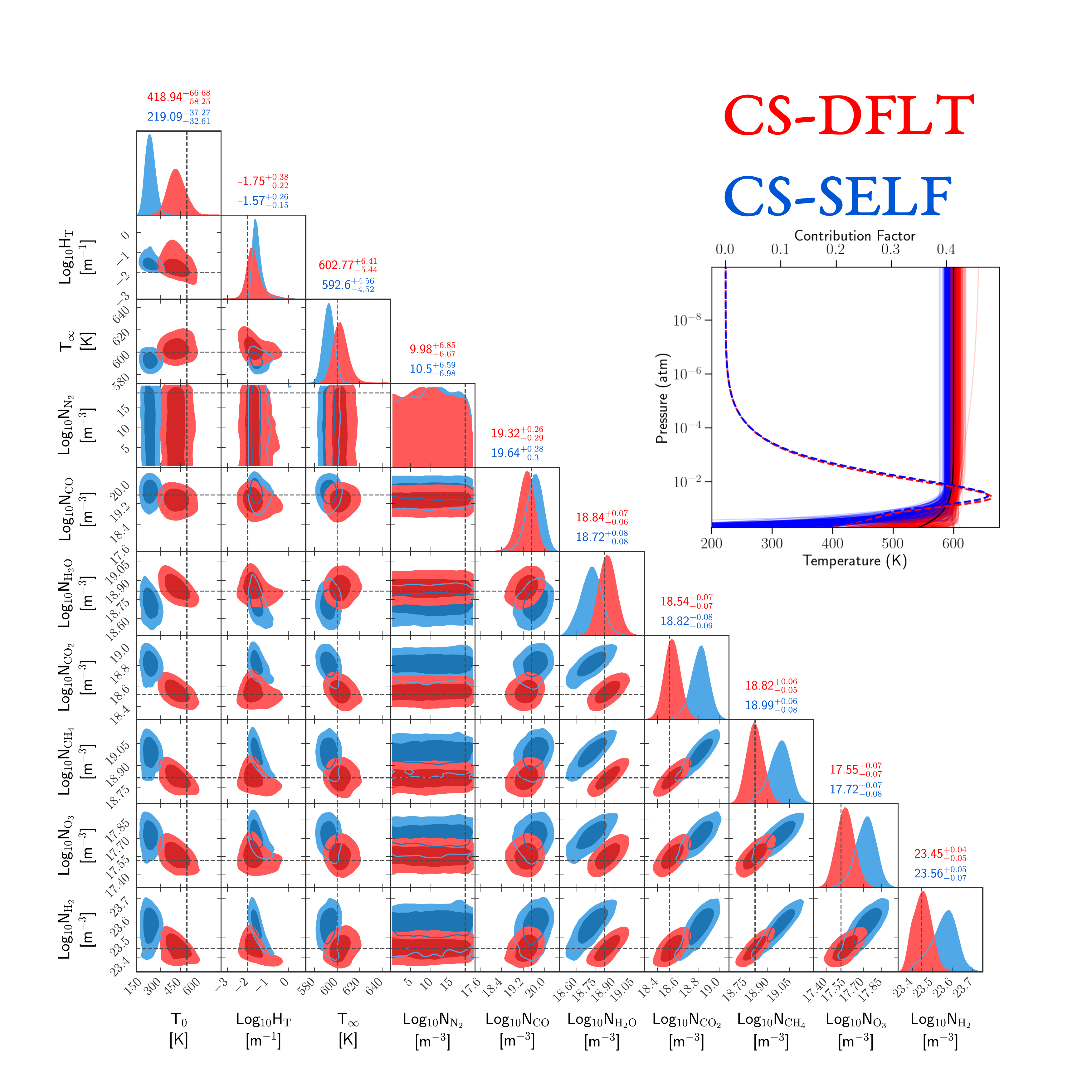}}
   \caption{\label{fig:HJ_CornerPlot} \textbf{Corner plot of the posterior probability distribution of the atmospheric parameters for a warm-Jupiter.} Corner plot showing the PPDs of the retrieved parameters for the case of the warm-Jupiter for self-retrieval (\texttt{CS-DFLT}, red) and cross-retrieval (\texttt{CS-SELF}, blue). The only difference between the two cross-sections relate to self-broadening; \texttt{CS-DFLT} assumes air-broadening (geocentric) while  \texttt{CS-SELF} assumes only self-broadening. Strong biases are seen in the retrieved value of T$_0$ (7.32$\sigma$), T$_\infty$ (2.64$\sigma$), carbon dioxide (7.32$\sigma$), methane (7.32$\sigma$), water (5.55$\sigma$), ozone (-5.57$\sigma$), and hydrogen (-3.97$\sigma$). (Top Right) 500 random PT profiles constructed from the posteriors are shown for comparison against the true profile shown in black. The dotted line shows the contribution factor can change significantly due to the changes in the pressure broadening values.}
\end{figure*}

\begin{figure*}[ht]
\centering
   {\vspace{-3cm}\includegraphics[width=0.95\textwidth, trim=0cm 0cm 0cm 0cm, clip=true]{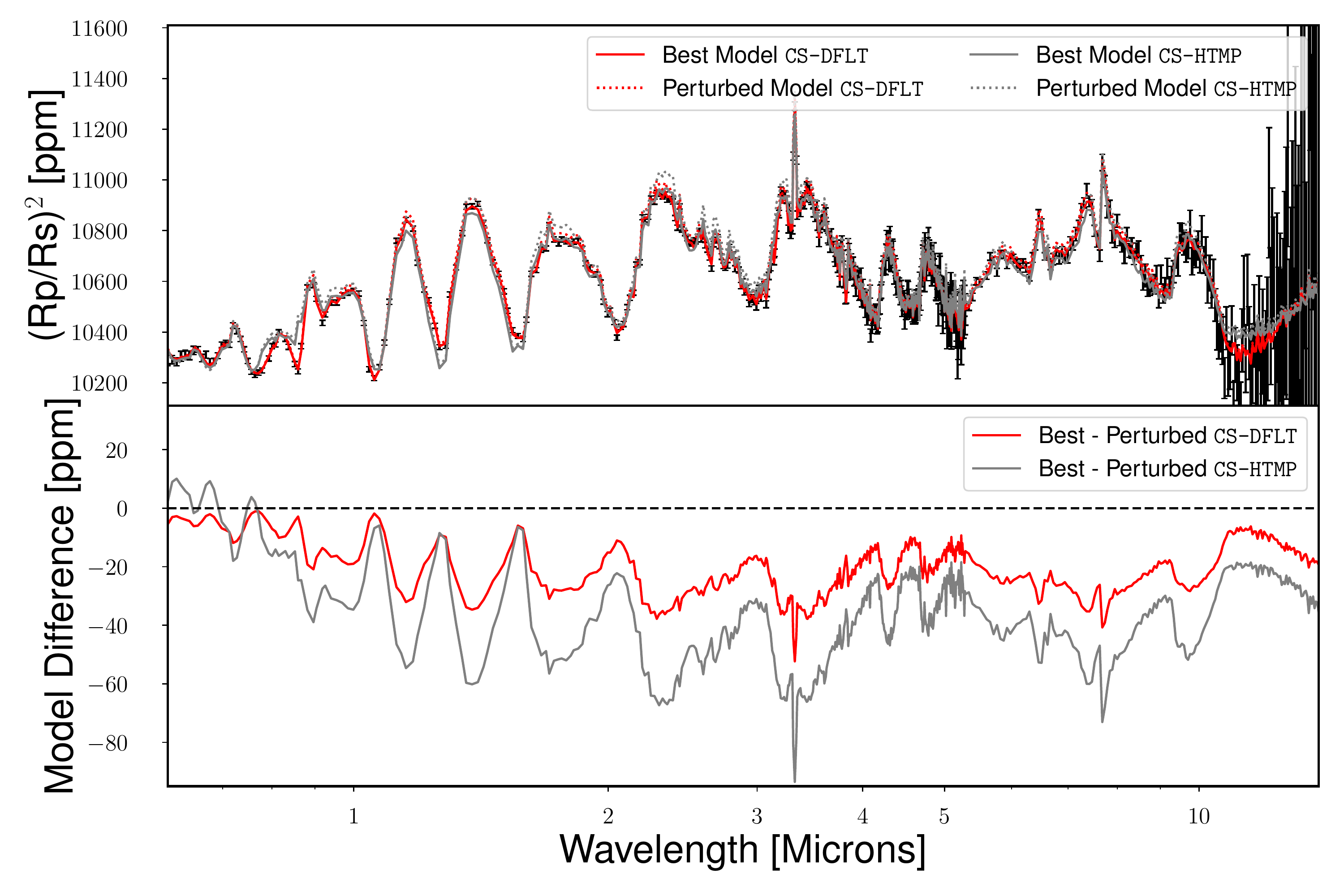}}
   \blue{\caption{\label{fig:CO_Perturbation} \textbf{Retrieval sensitivity to CO affected by opacity model.} Plot highlighting the difference in sensitivity to CO's number density between \texttt{CS-DFLT} (red) and \texttt{CS-HTMP} (gray) for the super-Earth case. (Top) Best-fit models are shown as solid lines while the models perturbed by -0.25 on Log$_{10}$N$_{\rm CO}$ are shown as dashed lines. (Bottom) Difference between best-fit and perturbed models, highlighting that the primary effect of a change in CO abundance is a change in scale height (all molecular features are affected). While a 0.25 change on Log$_{10}$N$_{\rm CO}$ from its best-fit value for \texttt{CS-DFLT} increases the $\chi^2$ by $\sim$1,000, it does increases the $\chi^2$ for \texttt{CS-HTMP} by $\sim$3,500 which explained the tighter constraint on CO reported in Figure~3 and Supplementary Table\,\ref{tab:CS_parameter}.}}
\end{figure*}

\end{methods}
\end{document}